\documentclass[11pt,letterpaper]{article}

\usepackage[letterpaper, top=0.5in, bottom=0.5in, left=0.5in, right=0.5in, footskip=.25in]{geometry}
\usepackage[utf8]{inputenc}
\usepackage[T1]{fontenc}
\usepackage{mathptmx} %Times new roman
\usepackage{amsmath}
\usepackage{amsfonts}
\usepackage{mathtools}
\usepackage{amssymb}
\usepackage{xcolor}
\usepackage[english]{babel}
\usepackage{tikz-network}
\usepackage{comment}
\usepackage{longtable}
\usepackage{multirow}
\usepackage{setspace} % Set linespacing

\usepackage{autobreak}
\allowdisplaybreaks

\usepackage{graphicx}
\graphicspath{ } %where the image is and called, or the following%
\usepackage[round]{natbib}   % omit 'round' option if you prefer square brackets
\usepackage[hyphens]{url}

\usepackage{bibspacing}
\usepackage{color}
\usepackage[hyphenbreaks]{breakurl}
\usepackage [colorlinks=true,breaklinks]{hyperref}%% needed for creating hyperlinks in the document, the option 
\definecolor{c1}{rgb}{0,0.4,0.7}%% r=red;g=green,b=blue
\definecolor{c2}{rgb}{0,0.4,0.7}%% light blue
\definecolor{c3}{rgb}{0.3,0,0.9}% red blue
\definecolor{c4}{rgb}{0,0.4,0.7}
\hypersetup{
	linkcolor={c1},% internal links
	citecolor={c2},% citations
	urlcolor={c4}  % external links/urls
}
\usepackage{thmtools}
\declaretheorem{theorem}
\declaretheorem{lemma}

\declaretheorem{definition}
\author{Xiaowei Hu}

\author{Xiaowei Hu}

\newcommand{\norm}[1]{\left\lVert#1\right\rVert}
\DeclarePairedDelimiter\abs{\lvert}{\rvert}%

\usepackage{authblk}
\author[a]{Xiaowei Hu\thanks{Corresponding author}}
\author[b]{Peng Li}
\affil[a]{Industrial and Manufacturing Engineering Department \protect\\ 
College of Engineering and Applied Science \protect\\ University of Wisconsin-Milwaukee \protect\\
3200 N Cramer St, Milwaukee, WI 53211, United States of America\vspace{.3cm}}

\affil[b]{Department of Supply Chain Management\protect\\ 
Rutgers Business School \protect\\ 
Rutgers University \protect\\ 
1 Washington Park, Newark, NJ 07102, United States of America\vspace{.3cm}}

{
    \makeatletter
    \renewcommand\AB@affilsepx{\protect\Affilfont}
    \makeatother

    \affil[ ]{}

    \makeatletter
    \renewcommand\AB@affilsepx{, \protect\Affilfont}
    \makeatother

    \affil[a]{hu8@uwalumni.com} 
    \affil[b]{peng.li.scm@rutgers.edu}
}

\title{Relief and Stimulus in a Cross-sector Multi-product \protect\\ Scarce Resource Supply Chain Network}

\date{\href{https://doi.org/10.1016/j.tre.2022.102932}{https://doi.org/10.1016/j.tre.2022.102932} \\
Received 25 May 2022; Revised 31 August 2022; Accepted 3 October 2022 \\
\textit{Transportation Research Part E: Logistics and Transportation Review} (2022), Volume 168, pp 102932.}

\begin{document}

\maketitle

\begin{flushleft}
\textbf{Abstract}
\end{flushleft}

In the era of a growing population, systemic changes to the world, and the rising risk of crises, humanity has been facing an unprecedented challenge of resource scarcity. Confronting and addressing the issues concerning the scarce resource's conservation, competition, and stimulation by grappling its characteristics and adopting viable policy instruments calls the decision-maker's attention with a paramount priority. In this paper, we develop the first general decentralized cross-sector supply chain network model that captures the unique features of scarce resources under a unifying fiscal policy scheme. We formulate the problem as a network equilibrium model with finite-dimensional variational inequality theories. We then characterize the network equilibrium with a set of classic theoretical properties, as well as with a set of properties that are novel to the network games application literature, namely, the lowest eigenvalue of the game Jacobian. Lastly, we provide a series of illustrative examples, including a medical glove supply network, to showcase how our model can be used to investigate the efficacy of the imposed policies in relieving supply chain distress and stimulating welfare. Our managerial insights inform and expand the political dialogues on fiscal policy design, public resource legislation, social welfare redistribution, and supply chain practice toward sustainability.

\begin{flushleft}
\textbf{Keywords}: networks; resource scarcity; game theory; supply chain management; fiscal policy 
\end{flushleft}

\textit{This research did not receive any specific grant from funding agencies in the public, commercial, or not-for-profit sectors.}

\newpage

\section{Introduction}
Humanity depends on the supply of an array of essential resources. These resources, such as agricultural crops, fisheries, wildlife, forests, petroleum, metals, minerals, even air, soil, and water, are not only critical to the flourishing of humanities but also considered assets for businesses operations, product innovation, and government affairs \citep{rosenberg1973innovative, Wagner2002, krautkraemer2005economics}. The impact of resource scarcity has also become a popular theme that impels some of the futuristic opinions and critiques on potential societal collapse and humanity trajectory (see also, \citet{friedman2009next, friedman2009hot}, \citet{gilding2012great}, inter alia). Therefore, meeting the resource demands is a substantive economic, commercial, and societal matter for the decision-makers.

However, modern history has never been immune to resource scarcity; in fact, we are experiencing shortages with greater magnitude and frequency. In the 1970s, the western world faced oil crises, which had a lasting impact on the United States and extended the oil shortage of 1979 to the farm crisis in the 1980s, resulting in low crop prices, foreclosed farmlands, and thereby the low farm income \citep{Corbett2013, federal1997examination}. Since the 2000s, many regions around the world, including North and South Africa, South and Central Asia, and some parts of the U.S., have been under enormous water stress. For instance, California of the U.S. declared to have experienced the driest year in 2013, and again in 2021 \citep{Worland2014, Jones2021}. Since 2019, induced by the COVID-19 global pandemic, the world has experienced an overwhelming shortage of commodities, ranging from raw materials such as lumber, precious metals, and labor, to the end-products such as medical supplies, semiconductor chips, grocery items, and household cleaning products. \citep{Gasparro2020, Chou2020, Valinsky2021}. As is summarized by an \citet{economist2021shortage} article: we are in a shortage economy. 

Some of the key causes of the resource scarcity, as is pointed out by Population Reference Bureau, include supply, demand, and structure \citep{ TheUnitedNations2012}. Today, with the world’s population projected to reach 8.5 billion in 2030 \citep{UnitedNations2019}, the demand for the planet's scarce resources continues to grow at a staggering rate. Meanwhile, the industrialization in the past two centuries has ended the “ever-cheaper” commodity resources era and set the fierce competition with rising prices for the “ever-scarcer” resources as the new norm \citep{krautkraemer1998nonrenewable}. Furthermore, traumatic crises, such as the aforementioned COVID-19 pandemic, also induce the scarcity of specific supplies.  

As a means to relieve resource shortages, stimulate growth, and influence economic outcomes, especially during critical times, supply-side fiscal policies are commonly used by governments as strategic instruments. The fiscal policy describes changes to government spending and revenue behavior \citep{keynes1936general}. In the energy crisis of the 1970s, the U.S. Senate advanced supply-side fiscal incentives to mitigate the ramification of the energy problems \citep{Long1973}. Those policies included investment credits for domestic exploratory drilling, research and development on the commercial exploitation of alternative energy sources. In the global financial crisis of 2007-2008, the U.S. Congress passed the \$787 billion American Recovery and Reinvestment Act in 2009, after the \$125 billion provided by the Economic Stimulus Act of 2008 \citep{davig2011monetary}. In the COVID-19 pandemic of 2020, worldwide governments adopted economic packages including fiscal, monetary, and financial policy measures to mitigate the negative effects of the public health crisis on the economy and to sustain public welfare \citep{gourinchas2020flattening}. To date, the estimated global stimulus effort has totaled \$ 10.4 trillion \citep{economist2021shortage} with \$5.2 trillion by the U.S. \citep{romer2021fiscal}. Monitoring the global fiscal policy in the wake of the COVID-19 pandemic, \citet{IMF2021} sees the worldwide fiscal stimulus in many advanced economies (e.g., the European Union and the U.S.) starting to make it more productive, equitable, and sustainable.

From the industrial standpoint of a scarce resource value chain, the supply-side fiscal policies, such as production incentives, tax credits, or expansionary funds, serve as a cost reduction and revenue increase for firms in short term and entrepreneurial processes in long term \citep{arestis2003reinventing, braunerhjelm2021rethinking}. In particular, the provision of fiscal policies to a resource sector in scarcity, directly or indirectly, can have an extreme preserve effect on the management of the targeted sector and the broader economy \citep{young2015fiscal}. For example, in the face of groundwater depletion in Indian states like Punjab and Gujarat, farmers are given free access to electricity to pump groundwater, in compensating for the massive underinvestment in water-saving technologies. Such a benefit results in a more profound impact on the resilience of the supply chain as well. Today, the emerging resource scarcity, especially from the latest pandemic, has served as a wake-up call by exposing the vulnerability of the supply chains. Along with an overwhelming pleading for fiscal stimulation in popular press  \citep{Blinder2021, Winegarden2021, Smith2021}, it bears merit to investigate how the supply-side fiscal policies could relieve the shortage and stimulate the social outcome. 

In existential resource scarcity where the demand is not limited to one single resource type, governments must identify their optimal grand strategies. But often, policies and regulations alone can inadvertently create sub-optimal signals to economic or environmental concerns. As a response to the need for expanding and informing political dialogues, growing acknowledgment of the cross-sector consideration such as the well-known water-energy-food (WEF) nexus concept \citep{hoff2011understanding} (see also, \citet{Bazilian2011} and the reference therein)
has emerged in the past decade. Similarly, the cross-sector construct considers key issues in each commodity sector's security through a holistic lens in order to achieve long-term sustainability (cf. \citet{biggs2015sustainable}). For instance, the discussions on integrated water resources management have now emphasized issues through analyses of inter-sectoral competition for surface freshwater resources, integration of water management at farm, system, and basin scales, or balancing tradeoffs with electricity generation and fuel supply \citep{Kurian2017, zhang2016energy}. To date, the cross-sectoral approach has tended toward technical assessments to enable knowledge and information sharing, support intersectoral cooperation, and enhance productivity and synergies \citep{biggs2015sustainable, olawuyi2020sustainable}.

In this paper, we construct a general supply chain network equilibrium (SCNE) model to better understand the scarce resources vis-à-vis conservation, competition, transportation, and policy intervention, by consolidating a resource's innate scarcity and the supply-demand caused shortage (more in \S\ref{subsec:SR}). Our supply chain network model incorporates the interdependency of the scarce resources by allowing for cross-resource sector links. With the inclusion of multiple resource products that are differentiated by markets, such a network emulates the scarce resource's heterogeneity feature in a similar approach by \citet{zhang2006network}. The interdependency and heterogeneity are two features inspired by \citet{pfeffer2003external}, and \citet{Hunt2000}. Furthermore, in the model, we install a unifying supply-side fiscal policy scheme in relation to the scarce resource ownership and production. We assess the efficacy of utilizing incentives and taxes in addressing the supply chain shortage from an industrial organization's viewpoint. Also, we incorporate a non-cooperative behavior of decision-makers, considering the unique ownership of the resources and the nature of quantity competition under the context of scarcity. We believe such behavior is best represented by a decentralized network, a construct known for its generality, efficiency, and ability to synthesize local information in supply chains (see also, \citet{lee1993material}, inter alia). Finally, in illustration, we provide a series of numerical examples, including one that concerns medical personal protective equipment (PPE) supplies, to illustrate the utility of our model. 

The remainder of this paper is organized as follows. Section 2 is a literature review and the contributions of our work. Section 3 presents the scarce resource supply chain network model with fiscal policies and its mathematical formulation. In Section 4, we show the solution concept of the model. Section 5 establishes the theoretical properties of the equilibrium in the context of both classic and novel network game studies. Section 6 proposes an algorithmic procedure to find the network equilibrium. Section 7 demonstrates the utility of the model with small-scale illustrative examples. Section 8 provides the decision-makers and practitioners with real-world managerial insights. Lastly, Section 9 is a highlight of our study and foresight to its extension. 

\section{Literature review}

\subsection{Scarce resources}\label{subsec:SR}

The broader studies on resource scarcity lie primarily in the fields of economics and management, each of which gives a self-containing viewpoint toward the characteristics of scarce resources. On one hand, the equilibrium and monopoly theories of Adam Smith, John Stuart Mill, and John Nash, as well as the rent and location theories of Alfred Marshall and Alfred Weber, tie the scarcity to the location and quality of resources needed for industrial and agricultural purposes \citep{weber1929alfred}. On the other hand, among the rising organizational theories, the \textit{Resource Dependence Theory} \citep{pfeffer2003external} claim the importance of resource by which the owners, producers, and suppliers are connected and interdependent (cf. \citet{Caniels2007}), whereas the \textit{Resource Advantage Theory} \citep{Hunt2000} posits resource’s features, including their demand heterogeneity and roles, in a firm setting. Overall, from an economic perspective, scarce resources can be characterized as of oligopolistic supply-side power, heterogeneous intra-sector demand, and low price-elasticity of products. From a management perspective, the possession of scarce resources is indicative of a firm’s competitive advantage and negotiation power. The coalescence of these two avenues of studies in resource scarcity has yet to occur in literature.

The scholarly interests in the interconnectedness of scarce resources started to agglomerate since the release of the World Economic Forum report introduced the WEF nexus as a novel concept \citep{hoff2011understanding}. Nearly all WEF-related literature that we are aware of has sought integrated solutions or suggested the interconnection of resources. For instance, \citet{bai2016enhanced} investigated the interplay between agricultural and fuel products; \citet{zhang2016energy} studied the interaction between water and energies; \citet{bakker2018shocks} exemplified a regional economy with water and agriculture in focus; \citet{mun2021designing} provided an expansionary solution to the hydro networks for energy, irrigation and flood control for developing countries. Almost all of these works, at the same time, amalgamate with agricultural supply chain management (SCM), in which resource scarcity was associated with a range of societal and economic issues. 

In the tumultuous times of critical resource shortages, especially in the wake of the COVID-19 pandemic, a wealth of studies have emerged to demonstrate the utility of quantitative supply chain models for the urgency of healthcare, agricultural, logistical, and humanitarian challenges. For example, \citet{kazaz2016interventions} developed a malaria medicine supply chain model to improve supply and reduce price volatility via interventions. \citet{arifouglu2022two} devised a decentralized system to combat the frequent supply and demand mismatches of the vaccines. Both \citet{chen2015economic} and \citet{liao2019information} used stylized models to examine the impact of information provision policies on farmers in developing countries. \citet{yu2020optimal} constructed three-echelon supply chain models to study donors' optimal subsidy strategies for the development of under-developed areas. \citet{chintapalli2021value} modeled, analyzed, and evaluated the impact of a credit-based scheme on the welfare of small-scale, risk-averse farmers. \citet{shen2021strengthening} explained the impact of the pandemic on supply chain resilience, identified the challenges that retail supply chains had experienced in China, and showcased the practical response of JD.com throughout the pandemic.

\subsection{Fiscal policies}\label{subsec:FP}

The research of fiscal policies such as economic incentives in behavioral economics is also related to our work. Under such subdomain, the design of fiscal policies often becomes a delicate and complex matter when presented with a decentralized system of multiple decision-making agents (cf. \citet{arrow2013public}). Generally, to assess the interplay of incentives and behaviors, microeconomic approaches are often adopted (see also, \citet{frey2001motivation}, \citet{huck2012social}, inter alia). With such approaches, incentives may be linked to the network topology and its design (cf. \citet{belhaj2013strategic}, \citet{jackson2015games}). For instance, \citet{calvo2004effects} showed that education subsidies and other labor market regulation policies display local increasing returns due to the network structure. \citet{hu2019socially} studied the price fluctuation in agricultural markets by discussing the role a carefully designed preseason buyout contract plays in individual farmer's welfare. \citet{srai2021interplay} explored how the interplay between competing and coexisting policy regimens can affect the supply dynamics between producers, customers, and their intermediaries in a supply network. 

As noted before, today's fragmentation and globalization of supply chains give impetus to new value-adding opportunities with the inclusion of supply-side fiscal policies. A plethora of recent works is dedicated to designing supply chain smart tax- and incentive-based schemes to promote new energies, foster sustainability, and develop e-commerce, health system, or broader economies. For example, \citet{alizamir2019analysis} analyzed the impact of two U.S. farmer's subsidy programs on consumers, farmers, and the government. \citet{jiang2021government} studied the government’s penalty provision in a bioenergy supply chain. \citet{tao2020network} evaluated the impact of various carbon policies on planning and operation decisions in emerging markets. \citet{xu2021supply} investigated the effect of the blockchain technology on the manufacturer’s operational decisions under cap-and-trade regulation. \citet{arifouglu2022two} developed a two-sided incentive program that proposes ``vaccination incentives'' to be given to both the demand and supply side to combat the frequent supply and demand mismatches. \citet{levi2022artificial} analyzed the effectiveness of various government interventions in improving consumer welfare under an artificial shortage in agricultural supply chains.

\subsection{Supply chain network equilibrium models}\label{subsec:SCNE}

Also akin to this study are the supply chain network models with multi-agent decision-making, in which case, game theory is often used (e.g., \citet{HuXiaowei2020}, inter alia). Among the wealth of existing literature, the SCNE models, inaugurated by \citet{Nagurney2002a}, permit one to represent the interactions between decision-makers in the supply chain for general commodities in terms of connections, flows, and prices. Since then, this sub-domain has proliferated to a wide range of applications with timely interests. In conjunction with resource scarcity and shortages, we note, for instance, \citet{wu2006modeling} and \citet{matsypura2007modeling} modeled power transmissions. \citet{masoumi2012supply}, \citet{dutta2019multitiered} each modeled the supply and demand of a unique medical item; \citet{besik2017quality} and \citet{wu2018perishable} modeled the logistics of fresh produce in the scope of their qualities and preservation. In conjunction with supply-side fiscal policies, on the other hand, we are aware that \citet{Yu2018}, \citet{wu2019closed}, and \citet{yang2021carbon, yang2022house}, for example, captured the design of an environmental tax or subsidy policy in a green supply chain. \citet{Nagurney2019a, nagurney2019strict} studied how global trade policies could impact the suppliers and markets. 

To date, albeit the research in SCNE models has been well developed, new studies continue to proliferate. In surveying the SCNE studies since 2020, we now identify a few trends of its application. First, the Chinese e-commerce industry has evolved rapidly with explosive consumer traffic in online shopping, payment, marketing, and services, as the works of \citet{zhang2020equilibrium}, \citet{guoyi2020research}, and \citet{chen2020pricing}, all affiliated with Chinese institutions, demonstrated the opportunities and utilities of such model in this industry. Second, the studies in socially responsible practices, led by firms' environmentally conscientious strategies, have reflected the zeitgeist of our renewed societal values. In particular, \citet{chen2020pricing}, \citet{zhang2021three}, and \citet{yang2021carbon, yang2022house} all showcased that corporate-level efforts like de-carbonization have embodied today's global supply chain practitioners' response. Last but not least, the COVID-19 pandemic has evidently created a host of complex problems in the logistics of medical supplies and labor, among which, for example, are the works by \citet{nagurney2021optimization, nagurney2021supply}, inter alia. 

\subsection{Research gap and current contribution}

While most of the literature in SCM focuses on either a category of general products or one specific type, it remains largely sparse on the interdependence of firms and the specific resource commodities they produce in a cross-sector framework (cf. \citet{Bell2012}). Such a gap can be exemplified by the papers referenced in \S\ref{subsec:SR}, as they mainly emphasize one or two specific products. To the best of our knowledge, we are not aware of a study in SCM that explores the characteristics of resource scarcity and the shortage conditions from both economic and management perspectives. A systematic review by \citet{Matopoulos2015} has underscored a need for further research on understanding the implications of resource scarcity for supply chain relationships and their impact on supply chain configurations.

Moreover, the fiscal policy scheme we implement in our model is inclusive of both incentives and taxes. We also set forth our scope of policy calibration as the broader social outcome and firm's collective profits, instead of specific performance measures (e.g., prices and quantities), as what appeared in the referenced papers in \S\ref{subsec:FP}. The coalescence of these two aspects has been limited in the theoretical literature on supply chain networks. Such a topic would have provided invaluable managerial insights and should, therefore, merit attention. The major contributions of this paper are summarized as follows: 

\begin{enumerate}

\item The construction of the first decentralized 4-tier multi-product scarce resource SCNE model with supply-side fiscal instruments. The model could handle not only the supply chains of scarce resources but also those of general resources that pertain to the aforementioned characteristics of scarcity. 

\item The first work to examine a general supply-side fiscal policy that encompasses both economic incentives and taxes in one model, with the social outcome as a calibration. 

\item A rigorous variational inequality formulation of the modeled general Nash equilibrium problem due to the consideration of both competition and resource capacities; The establishment of a set of novel theoretical conditions to characterize the network equilibria.
  		
\item Our examples and the resulted managerial insights instilled for governments and supply chain practitioners to handle scarce resource management, fiscal policy design, shortage relief, and sustainable development. 
  		
\end{enumerate} 

Finally, we use Table \ref{table:compare_SR} to compare and differentiate our work with other closely related SCNE models mentioned in this section. 

\begin{longtable}{  lccccc  p{15cm} } 
\caption{A comparison of closely related SCNE articles
\label{table:compare_SR}}\\
	\hline
\multirow{2}{*}{Article} & \multirow{2}{*}{Multi-product} & \multirow{2}{*}{Oligopolistic Competition} & \multirow{2}{*}{Capacitated} & \multicolumn{2}{c}{\underline{Policy Instrument}} \\
                         &                          &                        &                       & Single         & Unifying        \\\hline
\citet{masoumi2012supply}        &                          & \multicolumn{1}{c}{$\bullet$}                      &                       &                &                 \\
\citet{besik2017quality}         &                          & \multicolumn{1}{c}{$\bullet$}                      &                       &                &                 \\
\citet{Yu2018}                   &                          & \multicolumn{1}{c}{$\bullet$}                      &                       & \multicolumn{1}{c}{$\bullet$}              &                 \\
\citet{yang2021carbon}           &                          & \multicolumn{1}{c}{$\bullet$}                      &                       & \multicolumn{1}{c}{$\bullet$}              &                 \\
\citet{yang2022house}            &                          & \multicolumn{1}{c}{$\bullet$}                      &                       & \multicolumn{1}{c}{$\bullet$}              &                 \\
\citet{zhang2006network}         & \multicolumn{1}{c}{$\bullet$}                        &                        &                       &                &                 \\
\citet{nagurney2021multiclass}   & \multicolumn{1}{c}{$\bullet$} &                        &                       &                &                 \\
\citet{wu2006modeling}           &                          &                        &                       & \multicolumn{1}{c}{$\bullet$}              &                 \\
\citet{nagurney2019strict}       &                          &                        &                       & \multicolumn{1}{c}{$\bullet$}              &                 \\
\citet{wu2019closed}             &                          &                        &                       & \multicolumn{1}{c}{$\bullet$}              &                 \\
\citet{hu2013equilibrium}        &                          &                        & \multicolumn{1}{c}{$\bullet$} &                &                 \\
\citet{nagurney2017supply}       &                          &                        & \multicolumn{1}{c}{$\bullet$} &                &                 \\\hline
This Paper               & \multicolumn{1}{c}{$\bullet$} & \multicolumn{1}{c}{$\bullet$}                      & \multicolumn{1}{c}{$\bullet$} &                & \multicolumn{1}{c}{$\bullet$}   \\\hline
\end{longtable}

\section{The cross-sector multi-product scarce resource supply chain network model}

\begin{figure}[hbt!]
	\begin{center}
		\includegraphics[bb= 150 380 585 665, clip, width=0.8\textwidth]{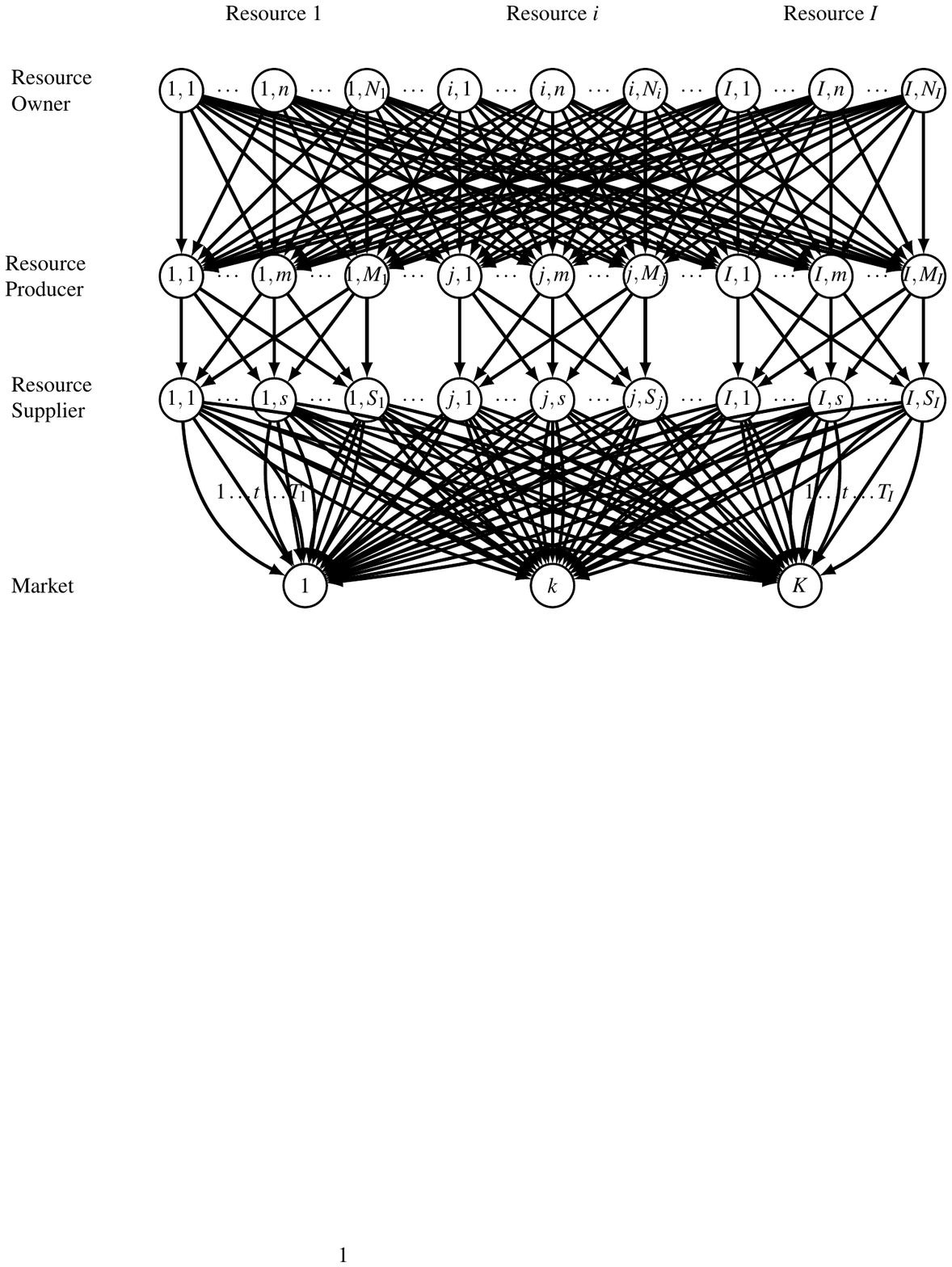} %left bottom right top
	\end{center} 
	\caption{A cross-sector scarce resource supply chain network model} \label{fig:Model}
\end{figure}

In the network, there are $I$ types of scarce resources, denoted as $i=1,...,I.$ We assign $i$'s alias $j$ in order to use it in different tiers. Naturally, $j=1,...,I$. Each type of resource is affiliated with $N_i$ owners, $M_j$ producers, and $S_j$ suppliers. The unfinished or end-product throughout the entire supply chain network is assumed to be homogeneous by the types of resources. See Figure \ref{fig:Model} for the network topology. Later, when referring to a certain type of resource, we will omit the phrase ``type'' or ``type of''.   

The first tier of nodes in the network represents the resource owner. For instance, a soybean-producing farmer is an owner of a resource type "soybean products". As the stakeholder of a scarce resource, an owner may supply such a resource to any producer. A typical resource owner is denoted as $(i,n)$, where $n=1,...,N_i$. 

The second tier of nodes represents the resource producer. An example of a resource producer can be a food manufacturer who uses the soybeans purchased from the farmers as raw material and produces soybean-related foods. A resource producer is able to obtain any type of resource from any owner, as stated above, but is only able to sell and ship its products to the suppliers who are specialized in the same type of resource. A typical resource producer is denoted as $(j,m)$, where $m=1,...,M_j$.  

The third tier of nodes represents the resource supplier. A resource supplier can only obtain and supply one type of resource to all markets. A soybean supplier, for instance, does not supply any other resources. A typical resource supplier is denoted as $(j,s)$, where $s=1,...,S_j$. 

The fourth tier of nodes represents the physical markets. A typical physical market, consuming some resources from its suppliers, is denoted as $k$, where $k=1,...,K$. Each supplier has $t$ transportation modes, where $t=1,...,T_j$, to ship a resource product to any physical market. To emulate the heterogeneity of the scarce resources characterized by \citet{Hunt2000}, we assume that each market has its own perception of the same resource, but is unable to distinguish the same resource from different suppliers or transportation modes. We assume also all markets to be competitive and, thereby, price-takers. 

We consider the network to be decentralized, i.e., each firm of a given tier competes with all others in a non-cooperative fashion to maximize its own profit by determining its optimal production quantity and shipments, given the simultaneous best response of all competitors. All of the notations regarding the model are listed in Table \ref{table:notations}.  

\begin{longtable}{  l  p{15cm} } 
\caption{Notations of the model\label{table:notations}}\\
	\hline
    Notation & Definition \\ \hline
    Indices: &   \\ 
    $i,j$ & The index of resource. $j$ is an alias of $i$. $i,j \in \{1,...,I \}$. \\ 
    $n$ & The index of resource owner $i$. $n \in \{1,...,N_i \}$.\\
    $m$ & The index of resource producer $j$. $m \in \{1,...,M_j \}$.\\
    $s$ & The index of resource supplier $j$. $s \in \{1,...,S_j \}$.\\
    $t$ & The index of transportation mode of resource product $j$. $t \in \{1,...,T_j \}$\\
    $k$ & The index of a physical market. $k \in \{1,...,K \}$.\\
    $g$ & The index of an incentive payment bracket. $g \in \{1,...,G \}$.\\
    \hline 
    
    Parameters: &   \\    
    $U_i$ & The total amount of resource $i$. $U \equiv \{U_1, U_2,..., U_I \}$ is a column vector.\\
    $A^i_g$ & The $g$'th incentive payment bracket of resource owners $(i,n)$\\
    $B^j_g$ & The $g$'th incentive payment bracket of resource producers $(j,m)$\\
    $\psi^{in}_{jm}$ & The non-negative conversion rate at resource producers $(j,m)$ from owner $(i,n)$\\
    $w_{ts}$ & The weight on market demand over supplier $s$ via transportation mode $t$ \\ \hline 
       
    Variables: &   \\ 
    $x^{in}_{jm}$ & The flow between resource owner $(i,n$) and producer $(j,m)$. \\ 
    $x^{jm}_{s}$ & The flow between resource producer $(j,m)$ and supplier $(j,s)$.\\
    $x^{js}_{tk}$ & The flow between resource supplier $(j,s)$ and market $k$ via transportation mode $t$. \\
	$x^{in},x_{jm}, x^{js}$ & The column vectors that collect $x^{in}_{jm}$, $x^{jm}_s$, $x^{js}_{tk}$, respectively.\\
	$d_{jk}$ & Demand of resource $j$ at market $k$. \\
	$p^{in}_{0jm}$ & The transaction price between resource owner $(i,n)$ to producer $(j,m)$. \\
	$p^{jm}_{1s}$ & The transaction price between resource producer $(j,m)$ to supplier $(j,s)$. \\
	$p^{js}_{2tk}$ & The sales price by supplier $(j,s)$ for market $k$ via transportation mode $t$. \\
	$p^{js}_{3tk}$ & The transaction price at market $k$ from supplier $(j,s)$ via transportation mode $t$. \\
    $\delta^{in}_g$ & Resource owner $(i,n)$'s excess of output quantity to bracket $g$. \\
    $\delta^{jm}_g$ & Resource producer $(j,m)$'s excess of output quantity to bracket $g$. \\
    $\lambda^0_i$ & The Lagrange multipliers associated with constraint (\ref{eqn:10}). \\
    $\lambda^1_{jm}$  & The Lagrange multipliers associated with constraint (\ref{eqn:30}). \\
    $\lambda^2_{js}$ & The Lagrange multiplier associated with constraints (\ref{eqn:35}). \\
    $\mu^0_{ing}$ & The Lagrange multipliers associated with constraint (\ref{eqn:11}). \\
    $\mu^1_{jmg}$  &  The Lagrange multipliers associated with constraint (\ref{eqn:31}). \\ 
    $\pi_{in}, \pi_{jm}, \pi_{js}$ & The profit of resource owner $(i,n)$, producer $(j,m)$, and supplier $(j,s)$, respectively. \\ 
    $CS_{jk}$  &  The consumer surplus of resource product $j$ at market $k$. \\ 
    $SW$  &  Social welfare of the entire supply chain network. \\ \hline	
        
    Functions: &   \\ 
    $\alpha^i_g(.)$ & The incentive payment to resource owners in bracket $g$. \\
    $\beta^j_g(.)$ & The incentive payment to resource producers in bracket $g$. \\
    $f^{in}= f^{in}(x^{in})$ & The operating cost incurred by resource owner $(i,n)$ in terms of outgoing shipments. \\
    $c^{in}_{jm}= c^{in}_{jm}(x^{in}_{jm})$ & The transaction cost  incurred by resource owner $(i,n)$ with producer $(j,m)$. \\
    $f^{jm}= f^{jm}(x_{jm})$ & The operating cost incurred by producer $(j,m)$ in terms of incoming shipments. \\
    $c^{jm}_{s}= c^{jm}_{s}(x^{jm}_{s})$ & The transaction cost incurred by producer $(j,m)$ with supplier $(j,s)$. \\
    $f^{js}= f^{js}(x^{js})$ &  The operating cost incurred by resource supplier $(j,s)$ in terms of outgoing shipments.\\
    $c^{js}_{tk}= c^{js}_{tk}(x^{js}_{tk})$ & The transaction cost incurred by supplier $(j,s)$ for market $k$ via transportation mode $t$.\\
    $\hat{c}^{js}_{tk}= \hat{c}^{js}_{tk}(x^{js}_{tk})$ & The transaction cost incurred by market $k$ with supplier $(j,s)$ via transportation mode $t$. \\
    $p^{j}_{3k} = p^{j}_{3k} (d_{jk})$ &  The price-demand function of resource $j$ at the market $k$. \\\hline     
\end{longtable}

\subsection{The fiscal policy scheme}

For the ease of conceptualization, in our model, we apply a supply-side fiscal incentive to resource owners and producers. We design a piece-wise scheme to represent some of the most commonly used governmental incentive policies. The incentive scheme devised here can easily be expanded to a general fiscal policy to include taxation. 

We now define the incentive scheme. Let $A^i_g$ denote the cutoff bracket of the incentive on resource $i$, with $g$ enumerating between $1,...G$. Let $\delta^{in}_g$ denote the resource owner $(i,n)$'s output quantity excess over the bracket $A^i_g$. We assume the incentives given to each entity to be a series of linear functions, only affected by its output quantity, as most commonly used governmental economic policies are (cf. \citet{sadka1976income, huck2012social, Yu2018}, inter alia). As such, we define the incentive payment function as
  \begin{equation}\label{eqn:owner_pay}
	\alpha^i_0(.)+\sum^{G}_{g=1} \alpha^i_g(.), 
  \end{equation}  
where $\alpha^i_0(.)$ is usually non-negative, both $\alpha^i_0(.)$ and $\alpha^i_g(.)$ are linear. For an arbitrary output quantity of $x$, the payment scheme for the resource owners is piece-wise linear in shape and can be expressed as:

If $0\leq x < A^i_1 $, then the incentive payment is $\alpha^i_0(x)$;

If $A^i_1 \leq x < A^i_2 $, then $\delta^{in}_1 =x - A^i_1$, and the incentive payment is $\alpha^i_0(x)+ \alpha^i_1(\delta^{in}_1)$;

If $A^i_2 \leq x < A^i_3 $, then $\delta^{in}_2 =x - A^i_2$, and the incentive payment is $\alpha^i_0(x) + \alpha^i_1(\delta^{in}_1) + \alpha^i_2(\delta^{in}_2)$;

If $A^i_g \leq x < A^i_{g+1} $, then $\delta^{in}_g =x - A^i_g$, and the incentive payment is $\alpha^i_0(x) + \alpha^i_1(\delta^{in}_1) + \alpha^i_2(\delta^{in}_2) +...+ \alpha^i_g(\delta^{in}_g)$;

If $x \geq A^i_G $, then the incentive payment is $\alpha^i_0(x) + \alpha^i_1(\delta^{in}_1) + \alpha^i_2(\delta^{in}_2) +...+ \alpha^i_g(\delta^{in}_G) = \alpha^i_0(x) + \sum^{G}_{g=1} \alpha^i_g(\delta^{in}_g)$.

Finally, a resource owner's general incentive payment function in terms of $\delta^{in}_g$ can be rewritten in a compact form as
  \begin{equation}\label{eqn:owner_scheme}
	\alpha^i_0(x) + \sum^{G}_{g=1} \alpha^i_g(\delta^{in}_g), \quad i=1,...,I,\quad  n=1,...,N, 
  \end{equation}
where,   
  \begin{equation}\label{eqn:owner_excess}
	\delta^{in}_g=max \{ x-A^i_g, \ 0 \}, \quad g=1,...,G.
  \end{equation}

In particular, when $\alpha^i_0(.)$ is set to be a constant and $\alpha^i_g(.)$ to be negative, expression (\ref{eqn:owner_scheme}) represents a regressive incentive.\footnote{A regressive incentive is one whose marginal rate decreases (cf. \citet{sadka1976income}).} An example of this incentive scheme is the United States Economic Impact Payments of COVID-19.\footnote{See: \href{https://www.irs.gov/coronavirus-tax-relief-and-economic-impact-payments}{https://www.irs.gov/coronavirus-tax-relief-and-economic-impact-payments}.} Naturally, with the properly chosen parameter values, expression (\ref{eqn:owner_scheme})-(\ref{eqn:owner_excess}) can also represent a general tax scheme. We will use this scheme to express the fiscal policies for the rest of the paper. Specifically, in \S\ref{subsec:producer}, we will invariably define the incentives administered to the resource producers.  

\subsection{The resource owner's problem}\label{subsec:owner}

A typical resource $i$ has a total capacity of $U_i$, shared by all $N_i$ owners of the same resource. A resource owner $(i, n)$ strategically determines its shipment quantity $x^{in}_{jm}$ to a resource producer $(j, m)$. We use $Q^0$, a column vector with $Q^0 \in R_{+}^{ \sum_{i=1}^{I}\sum_{j=1}^{I} N_i M_j }$, to group all  $x^{in}_{jm}$. We assume an owner incurs an operating cost of
  \begin{equation}\label{•}
	f^{in}= f^{in}(x^{in}), \ \forall i, n,
  \end{equation}
where $x^{in} \equiv (x^{in}_{11},...,x^{in}_{jm},...,x^{in}_{IM_j})$, and a transaction cost of 
  \begin{equation}\label{•}
	c^{in}_{jm}= c^{in}_{jm}(x^{in}_{jm}), \ \forall i, n,j,m.
  \end{equation}

The incentive payment received by the resource owner $(i, n)$ is 
  \begin{equation}\label{•}
	\alpha^i_0(\sum^{I}_{j=1} \sum^{M_j}_{m=1} x^{in}_{jm}) + \sum^{G}_{g=1} \alpha^i_g(\delta^{in}_g), \ \forall i, n,
  \end{equation}  
where, the quantity excess   
  \begin{equation}\label{eqn:excess_o}
	\delta^{in}_g=max \Bigg\{ \sum^{I}_{j=1} \sum^{M_j}_{m=1} x^{in}_{jm} - A^i_g,\ 0 \Bigg\}, \quad \forall g=1,...,G.
  \end{equation}
  
We group all $\delta^{in}_g$ into a column vector $\mathfrak{\Delta}^0$, with $\mathfrak{\Delta}^0 \in R_{+}^{\sum_{i=1}^{I} GN_i }$, and assume all of the cost functions above, i.e., $f^{in},\ c^{in}_{jm}$, to be continuously differentiable. 

The price of products between the resource owner $(i,n)$ and the producer $(j,m)$ is denoted by $p^{in}_{0jm}$. We further denote the equilibrium of such a price as $p^{in*}_{0jm}$. Hence, the total revenue of each owner is $\sum_{j=1}^{I} \sum_{m=1}^{M_j} p^{in}_{0jm} x^{in}_{jm}$.

As a profit-maximizer, each typical owner faces the following problem.
  \begin{align}
	\text{Maximize:} & \ \pi_{in}=\sum_{j=1}^{I} \sum_{m=1}^{M_j} p^{in}_{0jm} x^{in}_{jm} - f^{in}(x^{in})-\sum_{j=1}^{I} \sum_{m=1}^{M_j} c^{in}_{jm}(x^{in}_{jm}) + \alpha^i_0(\sum^{I}_{j=1} \sum^{M_j}_{m=1} x^{in}_{jm}) + \sum^{G}_{g=1} \alpha^i_g(\delta^{in}_g) && \label{eqn:ownerobj}\\
	\text{Subject to:} & \ \sum_{n=1}^{N_i} \sum_{j=1}^{I} \sum_{m=1}^{M_j} x^{in}_{jm} \leq U_i, && (\lambda^0_i) \label{eqn:10}\\
	& \ \sum^{I}_{j=1} \sum^{M_j}_{m=1} x^{in}_{jm} - \delta^{in}_g \leq A^i_g, \quad \forall g, && (\mu^0_{ing}) \label{eqn:11}\\
	& \ x^{in}_{jm} \geq 0, \ \forall j,m, \qquad \delta^{in}_g \geq 0, \ \forall g. &&\nonumber
  \end{align}

The constraint (\ref{eqn:10}) is the resource capacity. The constraint (\ref{eqn:11}), along with $\delta^{in}_g \geq 0 $, is the equivalence of expression (\ref{eqn:excess_o}). $\lambda^0_i$ and $\mu^0_{ing}$ are the Lagrange multipliers\footnote{For the details of such reformulation technique, see \citet{Bertsekas1989}.} associated with inequality constraints (\ref{eqn:10}) and (\ref{eqn:11}), respectively. We subsequently group all $\lambda^0_i$ and $\mu^0_{ing}$ into a column vector $\lambda^0$ and $\mu^0$, with $\lambda^0 \in R_{+}^I$ and $\mu^0 \in R_{+}^{\sum_{i=1}^{I} GN_i}$, respectively.
    
In this decentralized supply chain, the resource owners compete in a non-cooperative fashion, a behavior of classic Cournot-Nash \citep{Cournot2008, Nash1950, nash1951non} stating that each owner determines the optimal output, given the simultaneously optimal ones of all competitors. As such, the optimal behavior of all resource owners can be expressed as a variational inequality (VI)\footnote{\citet{gabaymoulin} were among some of the earliest landmark works to establish the equivalence between the optimization problem and variational inequality problem. See also, \citet{nagurney1999network}.}: determine $(Q^{0*},\mathfrak{\Delta}^{0*}, \lambda^{0*}, \mu^{0*}) \in \mathcal{K}^1$, satisfying:
  \begin{align}\label{eqn:owneropt2}
	\sum_{i=1}^{I} \sum_{n=1}^{N_i} \sum_{j=1}^{I} \sum_{m=1}^{M_j} \Bigg[ \frac{\partial f^{in}(x^{in*})}{\partial x^{in}_{jm}} + \frac{\partial c^{in}_{jm}(x^{in*}_{jm})}{\partial x^{in}_{jm}} - p^{in*}_{0jm} -  \frac{\partial \alpha^i_0(x^{in*}_{jm})}{\partial x^{in}_{jm}} + \lambda^{0*}_i + \sum_{g=1}^{G}\mu^{0*}_{ing} \Bigg] & \times (x^{in}_{jm} - x^{in*}_{jm})  \nonumber\\
	+ \sum_{i=1}^{I} \sum_{n=1}^{N_i} \sum_{g=1}^{G} \bigg[ - \frac{\partial \alpha^i_g(\delta^{in*}_g)}{\partial \delta^{in}_g} - \mu^{0*}_{ing} \bigg] \times (\delta^{in}_g - \delta^{in*}_g ) + \sum_{i=1}^{I} \bigg[U_i - \sum_{n=1}^{N_i} \sum_{j=1}^{I} \sum_{m=1}^{M_j} x^{in*}_{jm} \bigg] & \times  (\lambda^0_i - \lambda^{0*}_i)  \\ 	
	+ \sum_{i=1}^{I} \sum_{n=1}^{N_i} \sum_{g=1}^{G} \bigg[A^i_g - \sum^{I}_{j=1} \sum^{M_j}_{m=1} x^{in*}_{jm} + \delta^{in*}_g \bigg] & \times ( \mu^0_{ing} - \mu^{0*}_{ing}) \geq 0, \nonumber\\
	\forall (Q^0,\mathfrak{\Delta}^0, \lambda^0, \mu^0) \in \mathcal{K}^1, \nonumber
  \end{align} 
where, $\mathcal{K}^1 \equiv \lbrace (Q^0,\mathfrak{\Delta}^0, \lambda^0, \mu^0) \vert (Q^0,\mathfrak{\Delta}^0, \lambda^0, \mu^0) \in R_{+}^{\sum_{i=1}^{I} (\sum_{j=1}^{I} N_i M_j + 2G N_i) + I} \rbrace .$ 

Due to constraint (\ref{eqn:10}), the solution to the optimality condition (\ref{eqn:owneropt2}) is a variational equilibrium \citep{kulkarni2012variational}, a refinement of the general Nash equilibrium (GNE)\footnote{GNE problem has been frequently used in common-pool resource studies as a mathematical-economic tool. For more information on general Nash equilibrium problems (GNEP), see \citet{harker1991generalized}, \citet{Facchinei2007}.}. In a variational equilibrium, all of the Lagrange multipliers associated with the shared constraints are equal. By formulating a GNE in VI, we can resort to a set of well-established theoretical properties (more in Section \ref{sec:TheoP}) and algorithmic schemes than with a formulation by quasi-inequality, a formalism for GNE problems \citep{Facchinei2007}. 

In addition, condition (\ref{eqn:owneropt2}) proffers a readily interpretable mathematical form from an economic perspective. According to the first summand, a resource owner will ship a positive amount of the product to a resource producer if the price that the producer is willing to pay for the product is exactly equal to, the owner's marginal cost less the marginal incentive benefit. If the owner's marginal cost, less the marginal incentive benefit, exceeds what the retailer is willing to pay for the product, then there will not be any shipments from the owner to the producer. 

\subsection{The resource producer's problem}\label{subsec:producer}

A resource producer $(j,m)$ strategically determines its shipment quantity $x^{jm}_s$ to a resource supplier $(j, s)$. We group all $x^{jm}_{s}$ into a column vector $Q^1$,  with $Q^1 \in R_{+}^{\sum_{j=1}^{I} M_j S_j}$. We introduce a non-negative conversion rate of production $\psi^{in}_{jm}$. We assume that the producer's operating cost is associated with its total raw material quantity receivables from the owners and that it can be written as
  \begin{equation}\label{•}
	f^{jm}= f^{jm}(x_{jm}), \quad \forall j,m,
  \end{equation}
where, $x_{jm} \equiv (x_{jm}^{11},...,x_{jm}^{in},...,x_{jm}^{IN_i}).$

The producer's transaction cost is associated with the flow between each pair of producer and supplier 
  \begin{equation}\label{•}
	c^{jm}_{s}= c^{jm}_{s}(x^{jm}_{s}),\quad \forall j,m,s.
  \end{equation}  
  
Similar to the incentive scheme for the resource owners, we let $B^j_g$ denote the cutoff bracket of the incentives, with $g$ enumerating between $1,...G$. Let $\delta^{jm}_g$ denote the excess of output quantity over bracket $B^j_g$. As such, we define the resource producer's incentive payment function as 
  \begin{equation}\label{eqn:prod_pay}
	\beta^j_0(.)+\sum^{G}_{g=1} \beta^j_g(.), 
  \end{equation}
  
Specifically, the incentive payment received by the resource producer $(j,m)$ is a function of its total output quantity, i.e.,
  \begin{equation}\label{•}
	\beta^j_0(\sum^{S_j}_{s=1} x^{jm}_{s}) + \sum^{G}_{g=1} \beta^j_g(\delta^{jm}_g), \ \forall j, m,
  \end{equation}  
where, the quantity excess   
  \begin{equation}\label{eqn:excess_p}
	\delta^{jm}_g=max \Bigg\{\sum^{S_j}_{s=1} x^{jm}_{s}-B^j_g,\ 0 \Bigg\}, \quad \forall g=1,...,G.
  \end{equation}
    
We group all $\delta^{jm}_g$ into a column vector $\mathfrak{\Delta}^1$, and assume $f^{jm}$, $c^{jm}_{s}$ to be continuously differentiable, $\beta^j_0$, and $\beta^j_g$ to be linear. 

The price of products between the resource producer $(j,m)$ and the supplier $(j,s)$ is denoted by $p^{jm}_{1s}$. We further denote the equilibrium of such a price as $p^{jm*}_{1s}$. Hence, the total revenue of a producer is $\sum_{s=1}^{S_j} p^{jm}_{1s} x^{jm}_{s}$. 
		
As a profit-maximizer, each typical producer faces the following problem. 
  \begin{align}  
	\text{Maximize:} \ & \pi_{jm}=\sum_{s=1}^{S_j} p^{jm*}_{1s} x^{jm}_{s} - \sum_{j=1}^{I} \sum_{n=1}^{N_i} p^{in*}_{0jm} x^{in}_{jm} - f^{jm}(x_{jm}) - \sum_{s=1}^{S_j} c^{jm}_{s}(x^{jm}_{s}) + \beta^j_0(\sum^{S_j}_{s=1} x^{jm}_{s}) + \sum^{G}_{g=1} \beta^j_g(\delta^{jm}_g)  && \label{eqn:producerobj}\\ 
	\text{Subject to:} \ & \sum_{i=1}^{I} \sum_{n=1}^{N_i} x^{in}_{jm}\cdot\psi^{in}_{jm} \geq \sum_{s=1}^{S_j} x^{jm}_{s},  && (\lambda^1_{jm}) \label{eqn:30}\\
	& \sum^{S_j}_{s=1} x^{jm}_{s} - \delta^{jm}_g \leq B^j_g, \quad \forall g, && (\mu^1_{jmg}) \label{eqn:31}\\
	& x^{jm}_{s} \geq 0, \ \forall s, \qquad \delta^{jm}_g \geq 0, \ \forall g. && \nonumber
  \end{align}
  
The constraint (\ref{eqn:30}) is the flow conservation in light of the inter-resource conversion. The constraint (\ref{eqn:31}), along with $\delta^{jm}_g \geq 0 $, is the equivalence of expression (\ref{eqn:excess_p}). $\lambda^1_{jm}$ and $\mu^1_{jmg}$ are the Lagrange multipliers associated with inequality constraints (\ref{eqn:30}) and (\ref{eqn:31}), respectively. We subsequently group all $\lambda^1_{jm}$ and $\mu^1_{jmg}$ into a column vector $\lambda^1$ and $\mu^1$, with $\lambda^1 \in R_{+}^{\sum_{j=1}^{I} M_j}$ and $\mu^1 \in R_{+}^{\sum_{j=1}^{I} GM_j}$, respectively. The optimal behavior of all resource producers can be expressed as a VI: determine $(Q^{0*}, Q^{1*}, \mathfrak{\Delta}^{1*}, \lambda^{1*}, \mu^{1*}) \in \mathcal{K}^2$, satisfying:
  \begin{align}\label{eqn:produceropt2}
	\sum_{i=1}^{I} \sum_{n=1}^{N_i} \sum_{j=1}^{I} \sum_{m=1}^{M_j} \Bigg[ \frac{\partial f^{jm}(x^*_{jm})}{\partial x^{in}_{jm}} +  p^{in*}_{0jm} - \psi^{in}_{jm}\lambda^{1*}_{jm} \Bigg] & \times (x^{in}_{jm}-x^{in*}_{jm}) \nonumber\\
	+ \sum_{j=1}^{I} \sum_{m=1}^{M_j} \sum_{s=1}^{S_j}  \Bigg[ \frac{\partial c^{jm}_{s}(x^{jm*}_{s})}{\partial x^{jm}_{s}} - p^{jm*}_{1s} - \frac{\partial \beta^j_0(x^{jm*}_{s})}{\partial x^{jm}_{s}} + \lambda^{1*}_{jm} + \sum_{g=1}^{G} \mu^{1*}_{jmg} \Bigg] & \times (x^{jm}_{s} - x^{jm*}_{s}) \\
	+ \sum_{j=1}^{I} \sum_{m=1}^{M_j} \sum_{g=1}^{G} \bigg[ - \frac{\partial \beta^j_g(\delta^{jm*}_g)}{\partial \delta^{jm}_g} -\mu^{1*}_{jmg} \bigg] \times (\delta^{jm}_g - \delta^{jm*}_g ) + \sum_{j=1}^{I} \sum_{m=1}^{M_j} \sum_{g=1}^{G} \bigg[ B^j_g - \sum^{S_j}_{s=1} x^{jm*}_{s} + \delta^{jm*}_g \bigg] & \times (\mu^1_{jmg} - \mu^{1*}_{jmg}) \nonumber\\
	+ \sum_{j=1}^{I} \sum_{m=1}^{M_j} \bigg[ \sum_{i=1}^{I} \sum_{n=1}^{N_i} x^{in*}_{jm}\cdot\psi^{in}_{jm} - \sum_{s=1}^{S_j} x^{jm*}_{s} \bigg] & \times (\lambda^1_{jm} - \lambda^{1*}_{jm}) \geq 0, \nonumber\\
	\forall (Q^0, Q^1, \mathfrak{\Delta}^1, \lambda^1, \mu^1) \in \mathcal{K}^2, \nonumber
  \end{align}
where, $\mathcal{K}^2 \equiv \lbrace (Q^0, Q^1, \mathfrak{\Delta}^1, \lambda^1, \mu^1) \vert (Q^0, Q^1, \mathfrak{\Delta}^1, \lambda^1, \mu^1) \in R_{+}^{\sum_{j=1}^{I}(\sum_{i=1}^{I} N_i M_j + M_j S_j + 2GM_j
+ M_j)} \rbrace .$ 
  
\subsection{The resource supplier's problem}\label{subsec:supplier}

A typical resource supplier is denoted as $(j,s)$. Each supplier has $t$ modes to transport the resources to Market $k$. The supplier's strategic variable is $x^{js}_{tk}$, its shipment quantity to market $k$ with transportation mode $t$. We group all $x^{js}_{tk}$ into a column vector $Q^2$,  with $Q^2 \in R_{+}^{\sum_{j=1}^{I} KS_j T_j}$. We assume the supplier incurs an operating cost that is associated with the resource product flow from the producers, i.e., 
  \begin{equation}\label{•}
	f^{js}= f^{js}(x^{js}),
  \end{equation}
where, $x^{js} \equiv (x^{j1}_{s},...,x^{jm}_{s},...,x^{jM_j}_{s_j}), \ \ \forall j, s$. The supplier also incurs a transaction cost with each outgoing shipment to the markets: 
  \begin{equation}\label{•}
	c^{js}_{tk}= c^{js}_{tk}(x^{js}_{tk}).
  \end{equation}  
  
We assume $f^{js}$ and $c^{js}_{tk}$ to be continuously differentiable.

The price of goods between the resource supplier $(j,s)$ and the market $k$ via transportation mode $t$ is denoted by $p^{js}_{2tk}$. We further denote the equilibrium of such a price as $p^{js*}_{2tk}$. Hence, the total revenue of each supplier is $\sum_{t=1}^{T_j} \sum_{k=1}^{K} p^{js}_{2tk} x^{js}_{tk} $. 
	
As a profit-maximizer, each typical supplier faces the following problem.
  \begin{align}
	\text{Maximize:} & \quad \pi_{js}=\sum_{t=1}^{T_j}\sum_{k=1}^{K} p^{js*}_{2tk} x^{js}_{tk} - \sum_{m=1}^{M_j} p^{jm*}_{1s} x^{jm}_{s} - f^{js}(x^{js}) - \sum_{t=1}^{T_j}\sum_{k=1}^{K} c^{js}_{tk}(x^{js}_{tk}) & \label{eqn:supplierobj} \\
	\text{Subject to:} & \quad	\sum_{t=1}^{T_j} \sum_{k=1}^{K}x^{js}_{tk} \leq \sum_{m=1}^{M_j} x^{jm}_{s},  & (\lambda^2_{js}) \label{eqn:35} \\
	& \quad x^{js}_{tk} \geq 0,  \quad \forall t,k. & \nonumber
  \end{align}    

The constraint (\ref{eqn:35}) is the flow conservation of the finished resource commodities at each supplier, with $\lambda^2_{js}$ being the associated Lagrange multipliers. We subsequently group all $\lambda^2_{js}$ into a column vector $\lambda^2$ with $\lambda^2 \in R_{+}^{\sum_{j=1}^{I} S_j}$. The optimal behavior of all resource suppliers can be expressed as a VI: determine $(Q^{1*}, Q^{2*}, \lambda^{2*}) \in \mathcal{K}^3$, satisfying:
  \begin{align}\label{eqn:supplieropt2}
	\sum_{j=I}^{I} \sum_{s=1}^{S_j} \sum_{t=1}^{T_j} \sum_{k=1}^{K} \Bigg[\frac{\partial c^{js}_{tk}(x^{js*}_{tk})}{\partial x^{js}_{tk}} - p^{js*}_{2tk} + \lambda^{2*}_{js} \Bigg] & \times (x^{js}_{tk}-x^{js*}_{tk}) \nonumber \\
	+ \sum_{j=1}^{I} \sum_{m=1}^{M_j} \sum_{s=1}^{S_j} \bigg[ \frac{\partial f^{js}(x^{js*})}{\partial x^{jm}_{s}} + p^{jm*}_{1s} - \lambda^{2*}_{js} \bigg] & \times (x^{jm}_{s}-x^{jm*}_{s}) \\
	+ \sum_{j=I}^{I} \sum_{s=1}^{S_j} \Bigg[ \sum_{m=1}^{M_j} x^{jm*}_{s} - \sum_{k=1}^{K} \sum_{t=1}^{T_j} x^{js*}_{tk} \Bigg] & \times (\lambda^2_{js}-\lambda^{2*}_{js}) \geq 0, \qquad \forall (Q^1, Q^2, \lambda^2) \in \mathcal{K}^3, \nonumber
  \end{align} 
where, $\mathcal{K}^3 \equiv \lbrace (Q^1, Q^2, \lambda^2) \vert (Q^1, Q^2, \lambda^2) \in R_+^{\sum_{j=1}^{I} (KS_j T_j + M_j S_j + S_j)} \rbrace .$ 

\subsection{Demand market equilibrium conditions}

Scarce resources, such as clean water, energy, foods, and industrial materials, are consumed in demand markets with unique needs. Concerned with its willingness to pay, a market determines the purchasing quantity from each supplier via each available transportation mode. 

Because the difference in the suppliers or the transportation modes can cause different product preservation in the network, without loss of generality, for an arbitrary market $k$ ($k=1,...,K$) receiving resource commodity $j$ ($j=1,...,I$), we assign a set of ex-ante weights $w_{ts}$ across all flows via transportation modes $t$ ($t=1,...,T_j$) and suppliers $s$ ($s = 1,...,S_j $). In other words, the demand for product $j$ at market $k$ is
  \begin{equation}\label{eqn:45}
	d_{jk} = \sum_{t=1}^{T_j} \sum_{s=1}^{S_j} w_{ts} \cdot x^{js}_{tk}, \quad \forall j,k.
  \end{equation}   

We group all $d_{jk}$ from (\ref{eqn:45}) into a column vector $d$, i.e., $d \equiv (d_{11},...,d_{jk},...,d_{JK})$, where $d \in R^{IK}_+$. We further denote the equilibrium of such demands as $d^*_{jk}$ and $d^*$, respectively. Because each scarce-resource product in each market has a unique demand function, as is in (\ref{eqn:45}), it corresponds with a market price $p^{j}_{3k}$ in the form of price-demand function
  \begin{equation}\label{eqn:46}
	p^{j}_{3k} = p^{j}_{3k} (d) .
  \end{equation} 

Market $k$ incurs a transaction cost from supplier $(j,s)$ via transportation mode $t$ 
  \begin{equation}\label{•}
	\hat{c}^{js}_{tk}= \hat{c}^{js}_{tk}(x^{js}_{tk}), \quad \forall j,s,k,t.
  \end{equation}  
We assume $\hat{c}^{js}_{tk}$ to be continuously differentiable. Finally, we adopt the classic Spatial Price Equilibrium (SPE) model\footnote{The Spatial Price Equilibrium model appeared in the early literature of \citet{Samuelson1952}, \citet{takayama1964spatial}, inter alia, and was thereafter extended with VI theories by \citet{dafermos1984sensitivity}.} to represent the markets. The corresponding equilibrium condition is given by
  \begin{equation}\label{eqn:spe}
	p^{js*}_{2tk} + \hat{c}^{js*}_{tk}(x^{js*}_{tk}) \left\{
        \begin{array}{ll}
            = p^{j}_{3k} (d^*)  & \quad if \quad x^{js*}_{tk} > 0 \\
            \geq p^{j}_{3k} (d^*)  & \quad if \quad x^{js*}_{tk} = 0
        \end{array}
     \quad \forall j,s,t,k.
    \right.
  \end{equation} 

The intuition of equation (\ref{eqn:spe}) is, that the consumption of resource at the markets will remain positive if the supplier' purchase price of such a resource plus all associated cost is equal to the price consumers are willing to pay; however, if the supplier' purchase price plus cost turns out to be higher than what consumers are willing to pay, then there will be no shipments between the supplier and the market. 

As such, the optimal behaviors of all demand markets as a spatial price equilibrium can be expressed in a VI form: determine $(Q^{2*}, d^*) \in \mathcal{K}^4 $, satisfying:  
  \begin{align}\label{eqn:marketopt}
	\sum_{j=1}^{I} \sum_{s=1}^{S_j} \sum_{t=1}^{T_j} \sum_{k=1}^{K} [p^{js*}_{2tk} + \hat{c}^{js}_{tk}(x^{js*}_{tk})] \times (x^{js}_{tk}-x^{js*}_{tk}) - \sum^{I}_{j=1} \sum^{K}_{k=1} p^{j}_{3k} (d^*) \times (d_{jk} -d^*_{jk}) \geq 0, \qquad \forall (Q^2, d) \in \mathcal{K}^4, 
  \end{align} 
where, $\mathcal{K}^4\equiv \lbrace (Q^2, d) \vert Q^2 \in R_{+}^{\sum_{j=1}^{I} KS_j T_j},\ d \in R^{IK}_+  \rbrace .$  

\subsection{Welfare estimates}
To assess the social impact of the model elements in our supply chain network, we provide an estimate of consumer surplus and social welfare. The consumer surplus, termed to measure the aggregate of consumers' benefits upon purchasing a product (cf. \citet{willig1976consumer}) of resource $j$ at demand market $k$ is given by: 
	\begin{equation}\label{eqn:cs}
		CS_{jk} = \int_{0}^{d^*_{jk}} p^{j}_{3k}(z) dz - p^{j*}_{3k} d^*_{jk}, \quad \forall j,k.
	\end{equation}
	
We also denote social welfare of the network to be the aggregation of all firm profits and consumer surplus, i.e., 
	\begin{equation}\label{eqn:sw}
		SW = \sum_{i=1}^{I}\sum_{n=1}^{N_i} \pi_{in} + \sum_{j=1}^{I}\sum_{m=1}^{M_j} \pi_{jm} + \sum_{j=1}^{I}\sum_{s=1}^{S_j} \pi_{js} + \sum_{j=1}^{I}\sum_{k=1}^{K} CS_{jk}, \quad \forall i, j, n, m, s, k.	
	\end{equation}

\section{The network equilibrium }
	
\begin{definition}[Cross-sector multi-product scarce resource supply chain network equilibrium with fiscal policy] \label{def:def1}\hfill\\
A product flow pattern $(Q^{0*},Q^{1*},Q^{2*},\mathfrak{\Delta}^{0*}, \mathfrak{\Delta}^{1*}, d^*, \lambda^{0*}, \lambda^{1*}, \lambda^{2*}, \mu^{0*},\mu^{1*}) \in \mathcal{K}$ is said to constitute a cross-sector multi-product scarce resource supply chain network equilibrium with fiscal policy if it satisfies condition (\ref{eqn:owneropt2}), (\ref{eqn:produceropt2}), (\ref{eqn:supplieropt2}), and (\ref{eqn:marketopt}).
\end{definition}	
	
  \begin{theorem}[Variational inequality formulation] \label{thm:thm1}\hfill \\	
The equilibrium conditions governing the cross-sector multi-product scarce resource supply chain network, according to Definition 1, are equivalent to the solution to the following VIs: 

Determine $(Q^{0*},Q^{1*},Q^{2*}, \mathfrak{\Delta}^{0*}, \mathfrak{\Delta}^{1*}, d, \lambda^{0*}, \lambda^{1*}, \lambda^{2*}, \mu^{0*}, \mu^{1*})\in \mathcal{K}$, satisfying 
\begin{align}\label{eqn:60}
	\sum_{i=1}^{I} \sum_{n=1}^{N_i} \sum_{j=1}^{I} \sum_{m=1}^{M_j} \Bigg[ \frac{\partial f^{in}(x^{in*})}{\partial x^{in}_{jm}} + \frac{\partial f^{jm}(x^*_{jm})}{\partial x^{in}_{jm}} + \frac{\partial c^{in}_{jm}(x^{in*}_{jm})}{\partial x^{in}_{jm}} -  \frac{\partial \alpha^i_0(x^{in*}_{jm})}{\partial x^{in}_{jm}} + \lambda^{0*}_i - \psi^{in}_{jm}\lambda^{1*}_{jm}  + \sum_{g=1}^{G}\mu^{0*}_{ing} \Bigg] & \times (x^{in}_{jm} - x^{in*}_{jm}) \nonumber\\
	+ \sum_{j=1}^{I} \sum_{m=1}^{M_j} \sum_{s=1}^{S_j} \Bigg[ \dfrac{\partial f^{js}(x^{js*})}{\partial x^{jm}_{s}} +  \frac{\partial c^{jm}_{s}(x^{jm*}_{s})}{\partial x^{jm}_{s}} -  \frac{\partial \beta^j_0(x^{jm*}_{s})}{\partial x^{jm}_{s}} + \lambda^{1*}_{jm} - \lambda^{2*}_{js} + \sum_{g=1}^{G} \mu^{1*}_{jmg}  \Bigg] & \times (x^{jm}_{s} - x^{jm*}_{s}) \nonumber\\
	+ \sum_{j=I}^{I} \sum_{s=1}^{S_j} \sum_{t=1}^{T_j} \sum_{k=1}^{K} \Bigg[ \frac{\partial c^{js}_{tk}(x^{js*}_{tk})}{\partial x^{js}_{tk}} + \hat{c}^{js}_{tk}(x^{js*}_{tk}) + \lambda^{2*}_{js} \Bigg] & \times (x^{js}_{tk}-x^{js*}_{tk}) \nonumber\\
		+ \sum_{i=1}^{I} \sum_{n=1}^{N_i} \sum_{g=1}^{G} \bigg[ - \frac{\partial \alpha^i_g(\delta^{in*}_g)}{\partial \delta^{in}_g} - \mu^{0*}_{ing} \bigg] \times (\delta^{in}_g - \delta^{in*}_g ) + \sum_{j=1}^{I} \sum_{m=1}^{M_j} \sum_{g=1}^{G} \bigg[ -\frac{\partial \beta^j_g(\delta^{jm*}_g)}{\partial \delta^{jm}_g} - \mu^{1*}_{jmg} \bigg] & \times (\delta^{jm}_g - \delta^{jm*}_g ) \\
	- \sum^{I}_{j=1} \sum^{K}_{k=1} p^{j}_{3k} (d^*) \times (d_{jk} -d^*_{jk}) + \sum_{i=1}^{I} \bigg[U_i - \sum_{n=1}^{N_i} \sum_{j=1}^{I} \sum_{m=1}^{M_j} x^{in*}_{jm} \bigg] & \times  (\lambda^0_i - \lambda^{0*}_i) \nonumber \\ 
	+ \sum_{j=1}^{I} \sum_{m=1}^{M_j} \bigg[ \sum_{i=1}^{I} \sum_{n=1}^{N_i} x^{in*}_{jm}\cdot\psi^{in}_{jm} - \sum_{s=1}^{S_j} x^{jm*}_{s} \bigg] \times (\lambda^1_{jm} - \lambda^{1*}_{jm}) + \sum_{j=I}^{I} \sum_{s=1}^{S_j} \bigg[ \sum_{m=1}^{M_j} x^{jm*}_{s} - \sum_{k=1}^{K} \sum_{t=1}^{T_j} x^{js*}_{tk} \bigg] & \times (\lambda^2_{js}-\lambda^{2*}_{js}) \nonumber\\
	+ \sum_{i=1}^{I} \sum_{n=1}^{N_i} \sum_{g=1}^{G} \bigg[A^i_g - \sum^{I}_{j=1} \sum^{M_j}_{m=1} x^{in*}_{jm} + \delta^{in*}_g \bigg] \times ( \mu^0_{ing} - \mu^{0*}_{ing}) + \sum_{j=1}^{I} \sum_{m=1}^{M_j} \sum_{g=1}^{G} \bigg[ B^j_g - \sum^{S_j}_{s=1} x^{jm*}_{s} + \delta^{jm*}_g \bigg] & \times (\mu^1_{jmg} - \mu^{1*}_{jmg}) \geq 0, \nonumber \\
	\forall (Q^0,Q^1,Q^2, \mathfrak{\Delta}^0, \mathfrak{\Delta}^1, d, \lambda^0, \lambda^1, \lambda^2, \mu^0,\mu^1) \in \mathcal{K}, \nonumber
\end{align}  
where, $\mathcal{K} \equiv \mathcal{K}^1 \times \mathcal{K}^2 \times \mathcal{K}^3 \times \mathcal{K}^4 = \big\{ (Q^0,Q^1,Q^2, \mathfrak{\Delta}^0, \mathfrak{\Delta}^1, d, \lambda^0, \lambda^1, \lambda^2, \mu^0,\mu^1) \vert (Q^0,Q^1,Q^2, \mathfrak{\Delta}^0, \mathfrak{\Delta}^1, d, \lambda^0, \lambda^1, \\ \lambda^2, \mu^0,\mu^1) \in R_+^{\sum_{i=1}^{I} (\sum_{j=1}^{I} 2N_i M_j + 2G N_i) + \sum_{j=1}^{I} (2M_j S_j + 2KS_j T_j + 2GM_j + S_j + M_j) + IK+I} \big\}. $ 
  \end{theorem}

\begin{flushleft}
\textbf{Proof:} See Appendix \ref{appendix:a}.
\end{flushleft}
 
It should be noted that variable $p^{in*}_{0jm}$, $p^{jm*}_{1s}$, and $p^{js*}_{2tk}$ do not appear within the formulation of Theorem 1. They are endogenous to the model and can be retrieved once the solution is obtained. To retrieve $p^{js*}_{2tk}$, for all $j,s,t,k$, recall equilibrium condition (\ref{eqn:spe}). Since $p^{j}_{3k} (d^*)$ is readily available from (\ref{eqn:46}), if $x^{js*}_{tk}>0$ for some $j,s,t,k$, then $p^{js*}_{2tk}$ can be obtained by the equality 
	\begin{equation}\label{eqn:price2}
		p^{js*}_{2tk} = p^{j*}_{3k} (d^*) - \hat{c}^{js}_{tk}(x^{js*}_{tk}) . 
	\end{equation}

Invariably, if $x^{jm*}_{s}>0$ for some $j,m,s$, then from the second summand in (\ref{eqn:supplieropt2}), one may immediately obtain
	\begin{equation}\label{eqn:price1}
		p^{jm*}_{1s} = \lambda^{2*}_{js} - \frac{\partial f^{js}(x^{js*})}{\partial x^{jm}_{s}}. 
	\end{equation}
	
And, if $x^{in*}_{jm}>0$ for some $i,n,j,m$, then from the first summand of (\ref{eqn:produceropt2}),
	\begin{equation}\label{eqn:price0}
		p^{in*}_{0jm} = \psi^{in}_{jm}\lambda^{1*}_{jm} - \frac{\partial f^{jm}(x^*_{jm})}{\partial x^{in}_{jm}}. 
	\end{equation}

\section{Theoretical properties}\label{sec:TheoP}

We provide a few classic theoretical properties of the solution to VI (\ref{eqn:60}), based on \citet{gabaymoulin}, \citet{nagurney1999network}, and \citet{melo2018ving}, inter alia. In particular, we derive existence and uniqueness results by incorporating strategic measure of network games from the latest theoretical advancement. 

To facilitate the development in this section, we rewrite VI problem (\ref{eqn:60}) in standard form as follows: determine $X^* \in \mathcal{K}$ satisfying
  \begin{equation}\label{eqn:std}
	\langle F(X^*),\ X-X^* \rangle \geq0, \quad \forall X^* \in \mathcal{K},
  \end{equation}  
where, $X \equiv (Q^0,Q^1,Q^2, \mathfrak{\Delta}^0, \mathfrak{\Delta}^1, d, \lambda^0, \lambda^1, \lambda^2, \mu^0,\mu^1)$, with an indulgence of notation, $F(X) \equiv (F_{Q^0},F_{Q^1}, F_{Q^2}, F_{\mathfrak{\Delta}^0}, F_{\mathfrak{\Delta}^1}, F_{d}, F_{\lambda^0},\\ F_{\lambda^1}, F_{\lambda^2}, F_{\mu^0},F_{\mu^1})$, in which each component of $F$ is given by each respective summand expression in (\ref{eqn:60}), and, $\mathcal{K}\equiv \mathcal{K}^1 \times \mathcal{K}^2 \times \mathcal{K}^3 \times \mathcal{K}^4.$ The notation $\langle\cdot,\cdot\rangle$ represents the inner product in a Euclidean space. Both $F$ and $X$ are $\mathcal{N}$-dimensional column vectors, where $\mathcal{N} = \sum_{i=1}^{I} (\sum_{j=1}^{I} 2N_i M_j + 2G N_i) + \sum_{j=1}^{I} (2M_j S_j + 2KS_j T_j + 2GM_j + S_j + M_j) + IK+I $. 

First, we provide the existence properties. While $F$ in (\ref{eqn:std}) is continuous, the feasible set $\mathcal{K}$ is not compact. This causes the existence condition of (\ref{eqn:60}) not readily available. But one can impose a weak condition to guarantee the existence of a solution pattern, as in \citet{Nagurney2002a}. Let
  \begin{align}\label{eqn:bounds}
	\mathcal{K}_b \equiv \big\{ (Q^0,Q^1,Q^2, \mathfrak{\Delta}^0, \mathfrak{\Delta}^1, d, \lambda^0, \lambda^1, \lambda^2, \mu^0,\mu^1) \ \vert \  
	0 \leq Q^0 \leq b_1, 
	0 \leq Q^1 \leq b_2, 
	0 \leq Q^2 \leq b_3, 
	0 \leq \mathfrak{\Delta}^0 \leq b_4, \nonumber \\
	0 \leq \mathfrak{\Delta}^1 \leq b_5,
	0 \leq d \leq b_6,
	0 \leq \lambda^0 \leq b_7,  
	0 \leq \lambda^1 \leq b_8,
	0 \leq \lambda^2 \leq b_9,
	0 \leq \mu^0 \leq b_{10},
	0 \leq \mu^1 \leq  b_{11} \big\}, \nonumber
  \end{align} 
where $b=(b_1,b_2,b_3,b_4,b_5,b_6,b_7,b_8,b_9,b_{10},b_{11}) \geq 0$ and $Q^0 \leq b_1$, $Q^1 \leq b_2$, $Q^2 \leq b_3$, $\mathfrak{\Delta}^0 \leq b_4$, $\mathfrak{\Delta}^1 \leq b_5$, $d \leq b_6$, $\lambda^0 \leq b_7$, $\lambda^1 \leq b_8$, $\lambda^2 \leq b_9$, $\mu^0 \leq b_{10}$, $\mu^1 \leq  b_{11}$ means that $x^{in}_{jm} \leq b_1$, $x^{jm}_{s} \leq b_2$, $x^{js}_{tk} \leq b_3$, $\delta^{in}_g \leq b_4$, $\delta^{jm}_g \leq b_5$, $d_{jk} \leq b_6$, $\lambda^0_i \leq b_7$, $\lambda^1_{jm} \leq b_8$, $\lambda^2 \leq b_9$, $\mu^0_{ing} \leq b_{10}$, $\mu^1_{jmg} \leq  b_{11}$ for all $i,j,n,m,t,k$. Then $\mathcal{K}_b$ is a bounded, closed, convex subset of $R_+^{\mathcal{N}}$. Tangentially, the existence of $b$ is sensible from an economic perspective. Thus, the VI expression
  \begin{equation}\label{eqn:std_b}
	\langle F(X^b),\ X-X^b \rangle \geq0, \quad \forall X^b \in \mathcal{K}_b
  \end{equation}   
admits at least one solution. Following \citet{kinderlehrer2000introduction}, and \citet{Nagurney2002a}, it is straightforward to establish: 

  \begin{lemma}\label{lem:lem1}\hfill \\
The VI (\ref{eqn:std}) admits a solution if and only if there exists a $b>0$ such that the VI (\ref{eqn:std_b}) admits a solution in $\mathcal{K}_b$, with
  \begin{align}\label{eqn:bounds}
	Q^0 < b_1,\ 
	Q^1 < b_2,\ 
	Q^2 < b_3,\ 
	\mathfrak{\Delta}^0 < b_4,\ 
	\mathfrak{\Delta}^1 < b_5,\ 
	d < b_6, \\
	\lambda^0 < b_7,\ 
	\lambda^1 < b_8,\ 
	\lambda^2 < b_9,\ 
	\mu^0 < b_{10},\ 
	\mu^1 <  b_{11}. \nonumber
  \end{align}
  \end{lemma}

  \begin{theorem}[Existence]\label{thm:exist} \hfill \\
The VI problem (\ref{eqn:60}) admits at least one solution in $\mathcal{K}_b$. 
  \end{theorem}
\textbf{Proof:} 	
By virtue of theorem 3.1 in \citet{Harker1990a}, one can easily verify that $\mathcal{K}_b$ is non-empty, compact, and convex, and  that the mapping $F$ representing (\ref{eqn:60}) is continuous. Therefore, there exists a solution to the problem (\ref{eqn:60}) in $\mathcal{K}_b$. $\square$ 

Next, we provide a set of sufficient conditions for the uniqueness properties. In general, uniqueness is often associated with the monotonicity of the $F$ that enters the VI problem. Here, we begin by redefining a set of well-known monotonicities (definition 2.3.1 by \citet{facchinei2003finite}).

  \begin{definition}[Monotonicity]\label{def:mon}\hfill \\
A mapping $F: \mathcal{K} \subseteq R^n \rightarrow R^n$ is said to be \\
(a) monotone on $\mathcal{K}$ if  
    \begin{equation}\label{eqn:mon}
	\langle F(X^1) - F(X^2),\ X^1-X^2 \rangle \geq 0, \quad \forall X^1, X^2 \in \mathcal{K}.   
	\end{equation}
(b) strictly monotone on $\mathcal{K}$, if  
    \begin{align}\label{eqn:strictmon}
	\langle F(X^1) - F(X^2),\ X^1-X^2 \rangle > 0, \quad \forall X^1, X^2 \in \mathcal{K}, \quad X^1 \neq X^2.   
	\end{align}	
(c) strongly monotone on $\mathcal{K}$, if  
    \begin{equation}\label{eqn:strongmon}
	\langle F(X^1) - F(X^2),\ X^1-X^2 \rangle > \alpha \norm{X^1-X^2}^2, \quad \forall X^1, X^2 \in \mathcal{K},    
	\end{equation}
where $\alpha>0$.		
  \end{definition}
    
  \begin{definition}[Lipschitz Continuity]\label{def:lip}\hfill \\
$F(X)$ is Lipschitz Continuous on $\mathcal{K}$, if  
    \begin{equation}\label{eqn:lip}
	\langle F(X^1) - F(X^2),\ X^1-X^2 \rangle \geq L \norm{X^1-X^2}^2, \quad \forall X^1, X^2 \in \mathcal{K},   
	\end{equation}
where L is the Lipschitz constant.
  \end{definition}

We will also use the following established lemma (theorem 1.6 in \citet{nagurney1999network}). 
  \begin{lemma}\label{lma:unique_strict}\hfill \\
	Suppose that $F(X)$ is strictly monotone on $\mathcal{K}$, then the solution to the $VI(F,K)$ problem is unique, if one exists.  
  \end{lemma}
  
To further characterize the nature of the network, we need to adopt an additional set of notations. Following \citet{melo2018ving}, we define the game Jabocian\footnote{The game Jacobian is akin to the topology, equilibrium analyses, strategic interaction, and comparative statics of the network. For more details, see \citet{bramoulle2014strategic}, \citet{jackson2015games}, \citet{Parise2019}, \citet{melo2018ving}, and the reference therein.} of $F(X)$ as an $\mathcal{N}$-by-$\mathcal{N}$ matrix
    \begin{equation}\label{eqn:jacobian}
	J(X) \equiv [\nabla_q F_r(X)]_{q,r} = \Bigg[ \frac{\partial F_r(X)}{\partial X_q} \Bigg]_{q,r}, \quad \forall q,r=1,...,\mathcal{N},  X \in R_+^\mathcal{N}. 
	\end{equation}
  
Decomposing the Jacobian in terms of its diagonal and off-diagonal element yields
    \begin{equation}\label{eqn:jacdcmp}
	J(X) = D(X) + N(X),
	\end{equation}  
where $D(X)$ is a diagonal matrix, whose elements are
    \begin{equation}\label{eqn:diag}
	D_{qr}(X) = \left\{
        \begin{array}{ll}
            \frac{\partial F_r(X)}{\partial X_q} & \quad if\ q=r \\
            0 & \quad otherwise,
        \end{array}
    \right.
	\end{equation} 
and $N(X)$ is an off-diagonal matrix, whose elements are
    \begin{equation}\label{eqn:offdiag}
	N_{qr}(X) = \left\{
        \begin{array}{ll}
            \frac{\partial F_r(X)}{\partial X_q} & \quad if\ q \neq r \\
            0 & \quad otherwise.
        \end{array}
    \right.
	\end{equation}   
  
To accommodate the scenarios in which the Jacobian may be non-symmetric, we expand the definition of Jacobian in (\ref{eqn:jacdcmp}) by rewriting it as
    \begin{equation}\label{eqn:jacbardcmp}
	\bar{J}(X) = D(X) + \bar{N}(X),
	\end{equation}
where $\bar{N}(X) \equiv \frac{1}{2}[N(X)+N^T(X)]$.

Finally, we denote the lowest eigenvalue of a square matrix as $\lambda_{min}(\cdot)$.\footnote{For more information on how the lowest eigenvalue of a game Jacobian allows us to elicit insights on the interplay, neighboring influences, aggregate effect, see \citet{bramoulle2014strategic}.} Clearly, both $D(X)$ and $\bar{N}(X)$ have real-numbered eigenvalues because they are symmetric. Now we are ready to present the main results for the uniqueness. 

  \begin{theorem}[Sufficient conditions for Uniqueness]\label{thm:unique}\hfill \\	
	Assuming the condition for Theorem \ref{thm:exist} is satisfied, the solution $X^*$ to VI (\ref{eqn:60}) is unique, if
	
	(i) $F$ is strictly monotone on $\mathcal{K}$, or
	
	(ii) $F$ is strongly monotone on $\mathcal{K}$, or 
	
	(iii)	$\mathfrak{U}$ is strictly concave, and the condition $\abs{\lambda_{min}(\bar{N}(X))} < \lambda_{min}(D(X)) $ holds $\forall X \in \mathcal{K}$, where $\nabla_X \mathfrak{U}(X)= F(X)$. 
  \end{theorem}
  
\textbf{Proof:} 

   (i) The proof is immediate with Lemma \ref{lma:unique_strict} and Theorem \ref{thm:exist}. 
   
   (ii) See the proof of theorem 1.8 in \citet{nagurney1999network}.  
   
   (iii) Because all cost, incentive, and demand functions in equation (\ref{eqn:60}) are continuously differentiable, all of their partial derivatives are well defined. By virtue of proposition 4 in \citet{melo2018ving}, $F(X)$ is strictly monotone on $\mathcal{K}$. Combining Lemma \ref{lma:unique_strict} and Theorem \ref{thm:exist}, condition (iii) is proved.  $\square$

\textbf{Remark}

A few observations can be made. First, it is well-known that the monotonicity conditions in Definition \ref{def:mon} are ranked in the ascending order of "strength" \citep{facchinei2003finite}, i.e., (c) implies (b), and (b) implies (a). In Theorem \ref{thm:unique}, the uniqueness condition (i) and (iii), in essence, are established under strict monotonicity, whereas (ii) is under strong monotonicity. Hence, one can easily infer a relation of "(ii) $\Rightarrow$ (i)", as well as "(ii) $\Rightarrow$ (iii)" in Theorem \ref{thm:unique}.  

Second, to characterize the uniqueness of variational equilibrium (\ref{eqn:60}), one could also establish similar sufficient conditions by the semidefiniteness of $J(X)$ as it pertains to the analogous monotonicity features (cf. \citet{Parise2019, melo2018ving}).

\section{The algorithm}

To solve a VI problem in standard form, we propose an algorithm with theoretical measures of the result. The algorithm is extragradient method first proposed by \citet{korpelevich1976extragradient}, which is later promoted as the modified projection method in \citet{nagurney1999network} by setting the step length to 1. The solution is guaranteed to converge as long as the function $F$ that enters the standard form is monotone and Lipschitz continuous. The realization of the algorithm for the cross-sector multi-product scarce resource supply chain network model is given as follows. 
\begin{flushleft}
\textbf{The modified projection method:}
\end{flushleft}
Step 0. Initialization

Set $X^0 = (Q^{00},Q^{10},Q^{20}, \mathfrak{\Delta}^{00}, \mathfrak{\Delta}^{10}, d^0, \lambda^{00}, \lambda^{10}, \lambda^{20}, \mu^{00},\mu^{10}) \in \mathcal{K}$. Set $\tau=:1$ and select $\varphi$ such that $0 < \varphi \leq 1/L$, where $L$ is the Lipschitz constant for function $F$.   
\\
Step 1. Construction and computation 

Compute $\bar{X}^{\tau-1} = (\bar{Q}^{0 \tau}, \bar{Q}^{1\tau}, \bar{Q}^{2\tau}, \bar{\mathfrak{\Delta}}^{0\tau}, \bar{\mathfrak{\Delta}}^{1\tau}, \bar{d}^{\tau}, \bar{\lambda}^{0\tau}, \bar{\lambda}^{1\tau}, \bar{\lambda}^{2\tau}, \bar{\mu}^{0\tau},\bar{\mu}^{1\tau}) \in \mathcal{K}$ by solving the VI sub-problem
  \begin{equation}\label{eqn:compute}
	\langle \bar{X}^{\tau-1} + \varphi F(X^{\tau-1})-X^{\tau-1},\ X - \bar{X}^{\tau-1} \rangle \geq 0, \quad \forall X \in \mathcal{K}.
  \end{equation}
Step 2. Adaptation

Compute $X^{\tau} = (Q^{0\tau},Q^{1\tau},Q^{2\tau}, \mathfrak{\Delta}^{0\tau}, \mathfrak{\Delta}^{1\tau}, d^{\tau}, \lambda^{0\tau}, \lambda^{1\tau}, \lambda^{2\tau}, \mu^{0\tau},\mu^{1\tau}) \in \mathcal{K}$ by solving the VI sub-problem
  \begin{equation}\label{eqn:adapt}
	\langle X^{\tau} + \varphi F(\bar{X}^{\tau-1})-X^{\tau-1},\ X - X^{\tau} \rangle > 0, \quad \forall X \in \mathcal{K}.
  \end{equation}
Step 3. Convergence verification

If $\abs{X^{\tau}-X^{\tau-1}} \leq \epsilon$, for $\epsilon>0$, a pre-specified tolerance, then, stop; otherwise, set $\tau=:\tau+1$ and go to step 1.\\

The algorithm converges to the solution of $VI(F,\mathcal{K})$ under the following conditions.

  \begin{theorem}[Convergence]\label{thm:thm1}\hfill \\
		Assume that F(X) is monotone, as is in expression (\ref{eqn:strictmon}), and that $F(X)$ is also Lipschitz continuous, then the modified projection method converges to a solution of VI (\ref{eqn:std}).	
  \end{theorem}   
\textbf{Proof:} See Theorem 2.5 in \citet{nagurney1999network}.

\section{Small scale examples}

In this section, we construct several numerical cases to illustrate our model's utility. Example 1 is an application to the medical PPE glove supply. Example 2 broadens the application to an interconnected abstract resource-trio supply chain. The aforementioned algorithm is implemented in MATLAB installed on an ordinary ASUS VivoBook F510 personal laptop computer with an Intel Core i5 CPU 2.5 GHz and RAM 8.00 GB. We exhibit and discuss the highlights of each example. 

\begin{flushleft}
\textbf{Example 1.1: A medical gloves supply chain network (benchmark)}
\end{flushleft}

The COVID-19 pandemic of 2020 has reportedly caused a demand surge for medical gloves \citep{Finkenstadt2020}. In a single 100-day wave during early 2020, the estimated need for medical gloves was 3.939 billion \citep{Toner2020}, followed by a subsequent official guideline on conservation and optimizing usage of gloves during medical practice in the U.S. \citep{CDC2020}. It has become clear that the scarcity of medical gloves calls for a boost in the supply chain to the extent of better coordination and stimulus effort. Commonly used medical glove materials include latex, made from natural rubber, and nitrile, made from petroleum-based materials \citep{Anedda2020, Henneberry2020}. 

\begin{figure}[h]
		\centering
		\includegraphics[width=0.7\textwidth]{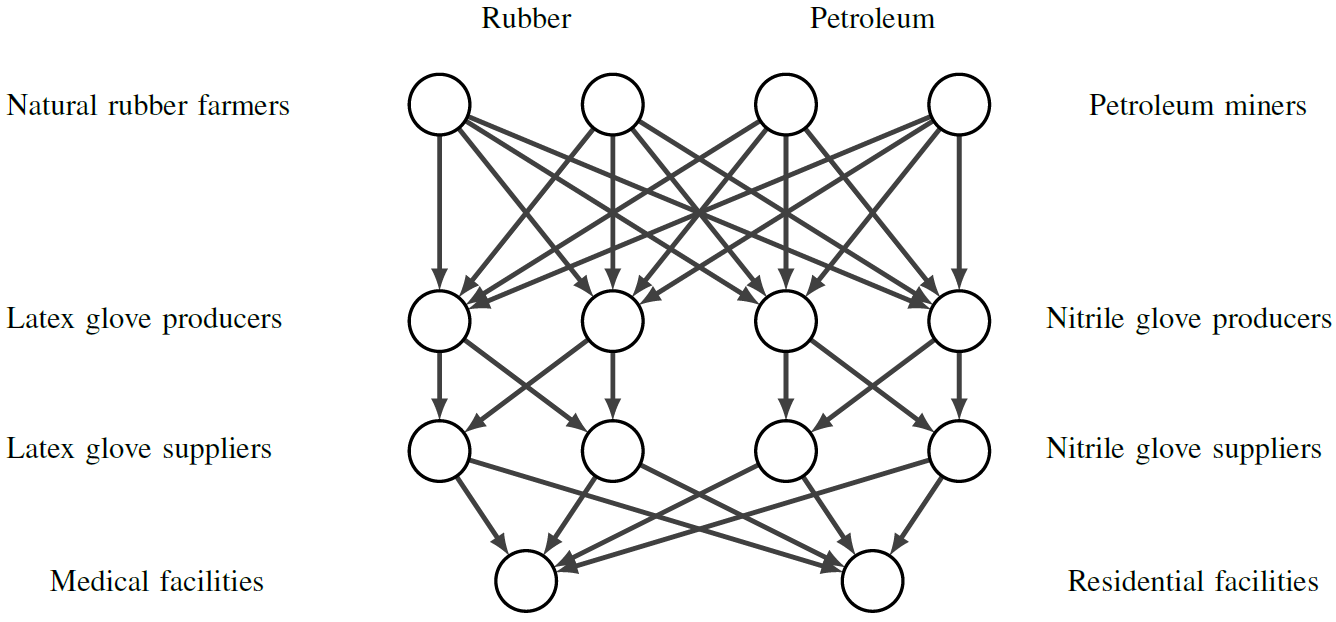}
		\caption{A resource-duo network with medical gloves supply chain}
		\label{fig:eg1}
	\end{figure}

This example illustrates a resource-duo supply chain network with natural rubber and petroleum as the resources, and medical gloves as their end-products. Specifically, the network contains 2 owners, 2 producers, and 2 retailers, as in Figure \ref{fig:eg1}. The corresponding end-products, latex and nitrile gloves, are shipped to in 2 demand markets, medical and residential facilities, via 1 available transportation mode. We use this example as our benchmark case, in which the supply chain network has unlimited resources and is imposed no fiscal policies. 
We will continue to use the same topology on example 1.2-1.5, with variations of setting. We examine the output quantities of firms, market prices, and welfare estimates. 

The cost functions are constructed for all $i=1,...,I$ and $j=1,...,J$, with $I=J=N_i=M_j=K_j=2$ and $T_j=1$, as the following. 
	\begin{align}
		&f^{1n}(x^{1n}) = 2.5 ( \sum^{I}_{j=1} \sum^{M_j}_{m=1} x^{1n}_{jm} )^2 + (\sum^{I}_{j=1} \sum^{M_j}_{m=1} x^{1n}_{jm})(\sum^{I}_{j=1} \sum^{M_j}_{m=1} x^{2n}_{jm}) + 2 \sum^{I}_{j=1} \sum^{M_j}_{m=1} x^{1n}_{jm}, \nonumber \\
		&f^{2n}(x^{2n}) = 0.5 ( \sum^{I}_{j=1} \sum^{M_j}_{m=1} x^{2n}_{jm} )^2 + (\sum^{I}_{j=1} \sum^{M_j}_{m=1} x^{1n}_{jm})(\sum^{I}_{j=1} \sum^{M_j}_{m=1} x^{2n}_{jm}) + 2 \sum^{I}_{j=1} \sum^{M_j}_{m=1} x^{2n}_{jm}, \nonumber \\
		&c^{jm}_{s}(x^{jm}_{s}) =0.5(x^{jm}_{s})^2, \qquad f^{js}(x^{js}) =0.1(\sum^{M_j}_{m=1} x^{jm}_s)^2, \qquad \hat{c}^{js}_{tk}(x^{js}_{tk}) =0.01x^{js}_{tk}, \nonumber \\
		&c^{11}_{11} = 0.5(x^{11}_{11})^2 + 3.5x^{11}_{11}, \quad c^{11}_{12}=0.5(x^{11}_{12})^2 + 3.5x^{11}_{12}, \quad c^{12}_{11} =0.5(x^{12}_{11})^2 + 2x^{12}_{11}, \quad c^{12}_{12}=0.5(x^{12}_{12})^2 + 2x^{12}_{12}, \nonumber \\
    	&c^{21}_{11} = 0.4(x^{21}_{11})^2 + 3.5x^{21}_{11}, \quad c^{21}_{12}=0.4(x^{21}_{12})^2 + 3.5x^{21}_{12}, \quad c^{22}_{11} =0.4(x^{22}_{11})^2 + 2x^{22}_{11}, \quad c^{22}_{12}=0.4(x^{22}_{12})^2 + 2x^{22}_{12}. \nonumber
	\end{align}
All other costs are set to zero. The price-demand functions at the markets are: 
	\begin{align}
		p^{j}_{3k} (d_{jk})= - d_{jk} + 300,\quad \forall j,k.\nonumber 
	\end{align}	

In addition, the conversion rates of production by resource producers are $\psi^{in}_{jm}=0.9$. The weights of resource commodities at the markets are $w_{11}=0.5, w_{12}=0.5$. 
The parameters concerning the resource capacities and policy instruments, $U_i$, $A^i_g$, $B^i_g$, are all set to a sufficiently large number, given their absence. 

We initialize the algorithm by setting all the flow quantity to be 1, the step-size $\varphi$ to be 0.01 (unless noted otherwise), the convergence tolerance $\epsilon$ to be $10^{-4}$. The computation process takes approximately 2.0 seconds. We display the solution in Table \ref{table:eg1}.\footnote{The complete result of all examples is available at: \href{https://github.com/Pergamono/SRSCN}{https://github.com/Pergamono/SRSCN}.} Among the equilibrium solution, the zero values of $\delta^{in}_g$ and $\delta^{jm}_g$ simply reaffirm the incentives to be flat-rate, whereas the zero values of all Lagrange multipliers suggest that the corresponding constraints are inactive. 

\begin{longtable}{  llllllll  p{13cm} } 
\caption{The equilibrium solution of example 1.1\label{table:eg1}}\\
	\hline
Variable & Result & Variable & Result & Variable & Result & Variable & Result \\ \hline
$x^{11*}_{11}$ & 8.88   & $x^{11*}_1$     & 13.79  & $\delta^{11*}_1$ & 0.00   & $\lambda^{0*}_1$    & 0.00   \\
$x^{11*}_{12}$ & 8.88   & $x^{11*}_2$     & 18.19  & $\delta^{12*}_1$ & 0.00   & $\lambda^{0*}_2$    & 0.00   \\
$x^{11*}_{21}$ & 9.50   & $x^{12*}_1$     & 13.79  & $\delta^{21*}_1$ & 0.00   & $\lambda^{1*}_{11}$ & 247.28 \\
$x^{11*}_{22}$ & 9.50   & $x^{12*}_2$     & 18.19  & $\delta^{22*}_1$ & 0.00   & $\lambda^{1*}_{12}$ & 247.28 \\
$x^{12*}_{11}$ & 8.88   & $x^{21*}_1$     & 14.53  & $\delta^{11*}_1$ & 0.00   & $\lambda^{1*}_{21}$ & 247.29 \\
$x^{12*}_{12}$ & 8.88   & $x^{21*}_2$     & 19.65  & $\delta^{12*}_1$ & 0.00   & $\lambda^{1*}_{22}$ & 247.29 \\
$x^{12*}_{21}$ & 9.50   & $x^{22*}_1$     & 14.53  & $\delta^{21*}_1$ & 0.00   & $\lambda^{2*}_{11}$ & 266.59 \\
$x^{12*}_{22}$ & 9.50   & $x^{22*}_2$     & 19.65  & $\delta^{22*}_1$ & 0.00   & $\lambda^{2*}_{12}$ & 263.64 \\
$x^{21*}_{11}$ & 8.88   & $x^{11*}_{11}$ & 13.79  & $\mu^{0*}_{111}$    & 0.00   & $\lambda^{2*}_{21}$ & 267.64 \\
$x^{21*}_{12}$ & 8.88   & $x^{11*}_{12}$ & 13.79  & $\mu^{0*}_{121}$    & 0.00   & $\lambda^{2*}_{22}$ & 264.97 \\
$x^{21*}_{21}$ & 9.50   & $x^{12*}_{11}$ & 18.18  & $\mu^{0*}_{211}$    & 0.00   & $d^*_{11}$         & 15.98  \\
$x^{21*}_{22}$ & 9.50   & $x^{12*}_{12}$ & 18.18  & $\mu^{0*}_{221}$    & 0.00   & $d^*_{12}$         & 15.98  \\
$x^{22*}_{11}$ & 8.88   & $x^{21*}_{11}$ & 14.53  & $\mu^{1*}_{111}$    & 0.00   & $d^*_{21}$         & 17.10  \\
$x^{22*}_{12}$ & 8.88   & $x^{21*}_{12}$ & 14.53  & $\mu^{1*}_{121}$    & 0.00   & $d^*_{22}$         & 17.10  \\
$x^{22*}_{21}$ & 9.50   & $x^{22*}_{11}$ & 19.67  & $\mu^{1*}_{211}$    & 0.00   &                  &        \\
$x^{22*}_{22}$ & 9.50   & $x^{22*}_{12}$ & 19.67  & $\mu^{1*}_{221}$    & 0.00   &                  &       \\
\hline     
\end{longtable}

\begin{flushleft}
\textbf{Example 1.2: The commons\footnote{A commons is where the natural resources are owned and shared collectively \citep{ostrom1990governing}.} with a resource capacity limit}
\end{flushleft}

The natural rubber shipments from the commons of farming and harvesting, which are concentrated in Southeast Asia, can be severely affected by external shocks, such as natural disasters, geopolitical shifts, regulations, and pandemics \citep{Chou2020, Lee2020}. Therefore, it bears merit to investigate the resilience of these supply chains. We perform a sensitivity study on the rubber capacity limit, by inheriting all settings from example 1.1, with the additional imposition of resource capacity on natural rubber. Acknowledging $U_1$'s level when such a limit can be reached, we vary $U_1$ between 10 and 100. 
\begin{figure}[h!]
    \centering
    \begin{minipage}{0.45\textwidth}
        \centering
        \includegraphics[width=1\textwidth]{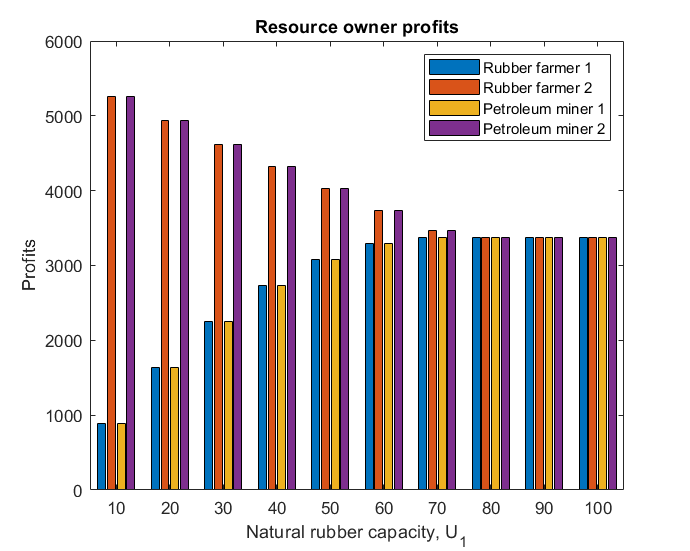} % first figure itself
        \caption{Owner profits under a resource capacity}
        \label{fig:eg2_1}
    \end{minipage}\hfill
    \begin{minipage}{0.45\textwidth}
        \centering
        \includegraphics[width=1\textwidth]{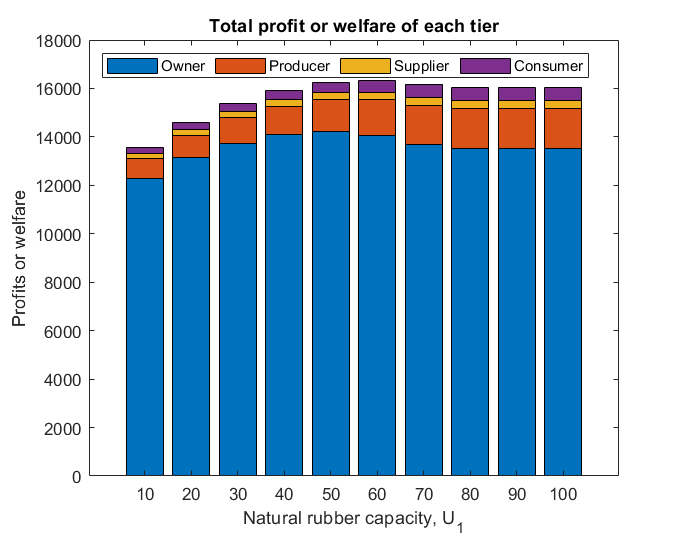} % second figure itself
        \caption{Total profits and welfare under a resource capacity}
        \label{fig:eg2_2}
    \end{minipage}
\end{figure}
As the natural rubber's capacity increases, as expected, each glove producer, supplier's profit, as well as the consumer surplus, increases. The petroleum miners, however, suffer from the abundance of natural rubber, as Figure \ref{fig:eg2_1} shows. The total profits or surplus of each supply chain tier are displayed in Figure \ref{fig:eg2_2}. It is worth noting that the overall height of each stacked bar indicates social welfare of the entire network. Interestingly, we find that social welfare peaks at a $U_1$'s level of around 60, likely owing to a similar trend of the owners' total profit that peaks at the similar level of $U_1$. 

\begin{flushleft}
\textbf{Example 1.3: Who should get the incentive, owners or producers?}
\end{flushleft}

In this example, we compare the scenarios in which either the natural rubber farmers or the latex glove producers receive a fairly small flat-rate incentive on their production quantity. As such, we inherit all settings from example 1.1, with the additional imposition of a fiscal policy in the form of quantity incentives. Specifically, we separately incentivize the natural rubber farmers and the latex glove producers with a flat-rate of $\alpha^1_0$ and $\beta^1_0$, respectively. 
\begin{figure}[h!]
    \centering
    \begin{minipage}{0.48\textwidth}
        \centering
        \includegraphics[width=0.9\textwidth]{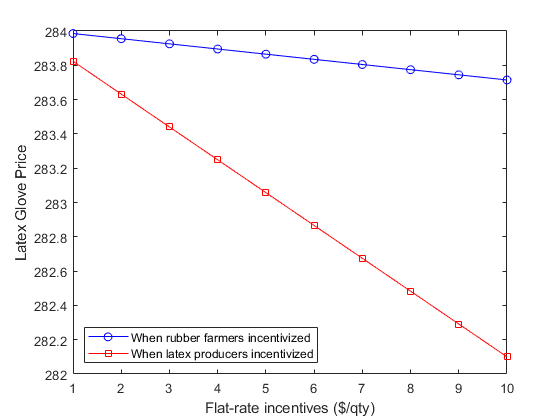} 
        \begin{center}
        (a) Latex glove market price
        \end{center}        
    \end{minipage}\hfill
    \begin{minipage}{0.48\textwidth}
        \centering
        \includegraphics[width=0.9\textwidth]{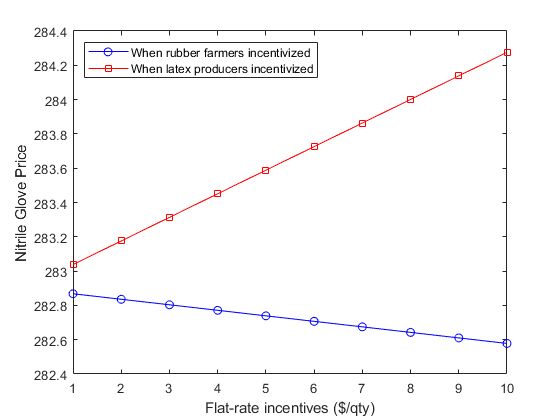} 
        \begin{center}
        (b) Nitrile glove market price
        \end{center} 
    \end{minipage}
    \caption{Glove market prices by flat-rate incentive}
    \label{fig:eg3_p}
\end{figure}

We elect to examine only the glove prices at the markets, as the rest of the equilibrium results can be examined invariably. In Figure \ref{fig:eg3_p}, we observe that with the flat-rate incentives administered to the latex glove supply chain, the corresponding latex glove prices at the market are reduced, whereas the prices of the substitute product, nitrile glove, change in opposite trends. 
\begin{figure}[h!]
		\centering
		\includegraphics[width=\textwidth]{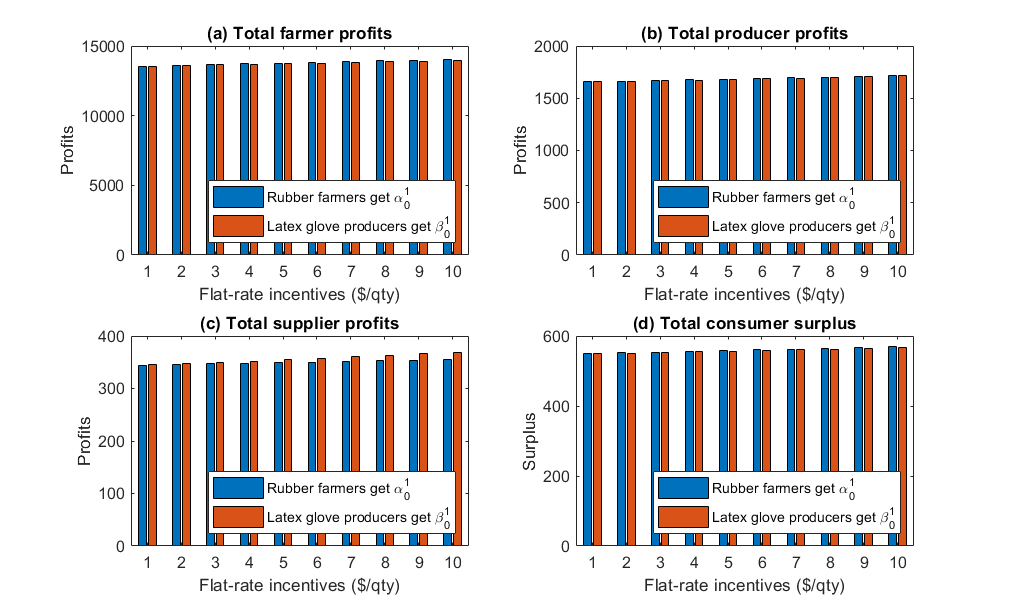}
		\caption{Total profits and welfare under the flat-rate incentives}
		\label{fig:eg3_1}
	\end{figure}
	\begin{figure}[h!]
		\centering
		\includegraphics[width=13cm]{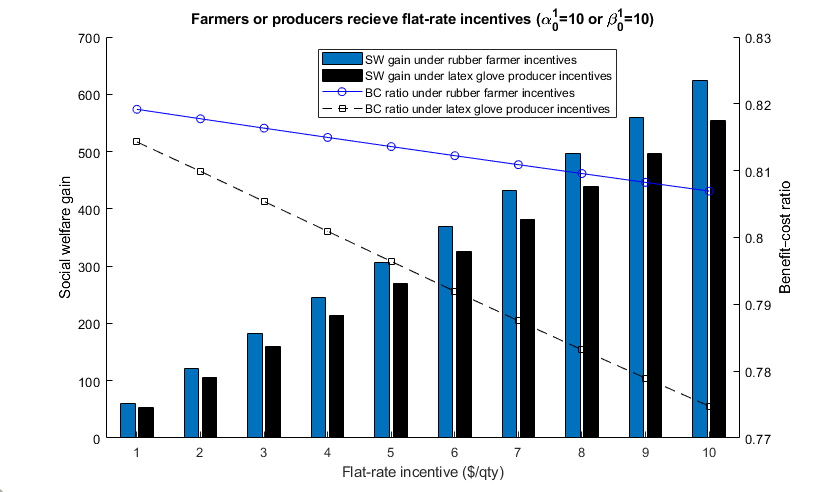}
		\caption{Social-welfare gains and policy efficiency under the flat-rate incentives}
		\label{fig:eg3_2}
	\end{figure}
	
From the standpoint of the supply chain participants, the farmers, producers, and consumers each as a tier, benefit mildly, because of the relatively small amount of incentives. See Figure \ref{fig:eg3_1}. The suppliers, however, enjoy a discernible gain in total profit when the incentives are given to the rubber farmers. 

From the standpoint of the incentive administer, it bears meaning to examine such a fiscal policy's social benefit and efficiency. We use the Benefit-Cost ratio, i.e., the dollar amount of social-welfare gain for every \$1 incentive administered to the system, to represent the efficiency. In Figure \ref{fig:eg3_2}, we show the social-welfare gains, as well as the efficiency of the incentive. It can be seen that the efficiency of the incentive given to the resource producer suffers more severely.

\begin{flushleft}
\textbf{Example 1.4: Regressive vs. flat-rate incentives}
\end{flushleft}

In this example, we examine how a regressive incentive policy differs from a flat-rate one, as well as how the incentive bracket affects the system performance. Once again, we inherit all settings from example 1.1, plus the additional imposition of regressive incentive policy, with $\alpha^1_0=11$ and $\alpha^1_1=-2.2$ on both natural rubber farmers, to be selected to provide a sensible comparison against a flat-rate, $\alpha^1_0=10$, incentive policy. The only incentive bracket, $A^1_1$, will be left varying as a sensitivity study parameter.

\begin{figure}[h!]
    \centering
    \begin{minipage}{0.475\textwidth}
        \centering
        \includegraphics[width=1\textwidth]{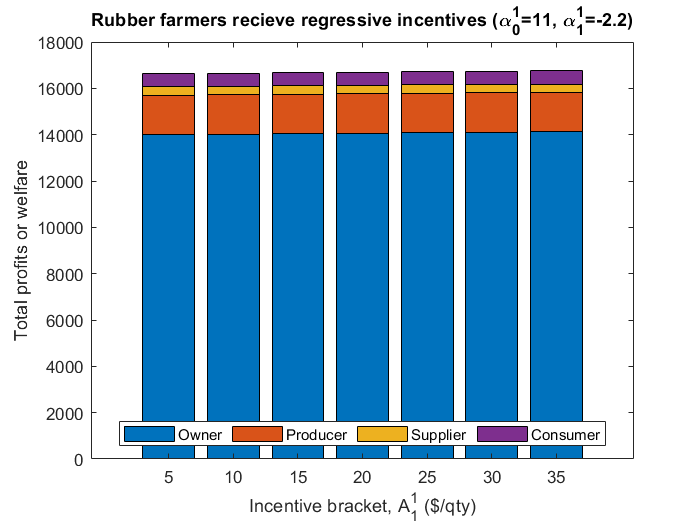} % first figure itself
        \caption{Total profits and welfare under the regressive incentives}
        \label{fig:eg4_1}
    \end{minipage}\hfill
    \begin{minipage}{0.475\textwidth}
        \centering
        \includegraphics[width=1\textwidth]{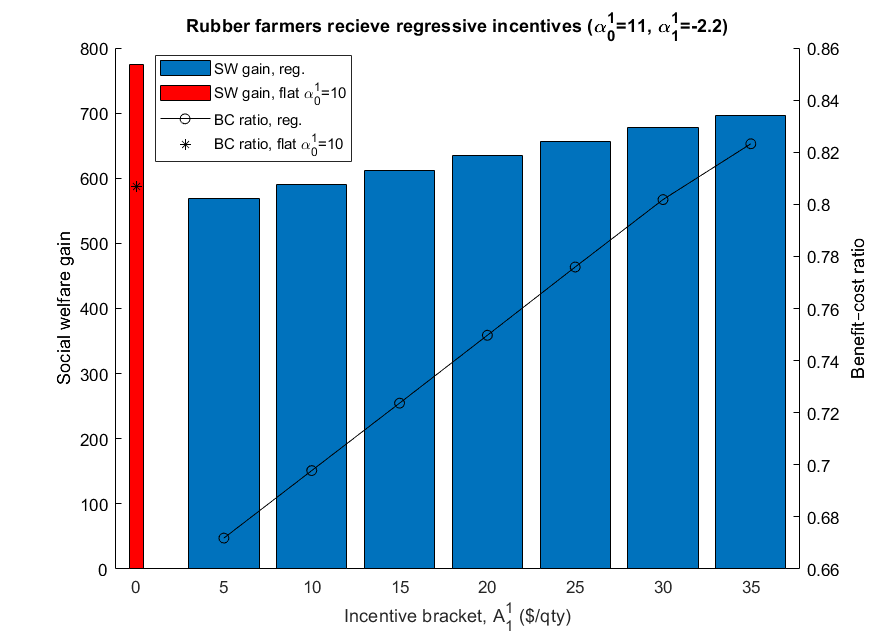} % second figure itself
        \caption{Social-welfare gains and Benefit-Cost ratio under the regressive incentives}
        \label{fig:eg4_2}
    \end{minipage}
\end{figure}

The range of $A^1_1$ displayed in Figure \ref{fig:eg4_1} and \ref{fig:eg4_2} are selected to ensure each farmer's total output quantity exceeds such a bracket value for it to take effect. In Figure \ref{fig:eg4_1}, we observe that the change of the regressive incentive bracket does not influence each supply chain tier's profit and the consumer welfare significantly. Such a result is consistent with the findings with respect to the effect of the ``tax brackets'' on firm's profits in a study by \citet{Yu2018}. In Figure \ref{fig:eg4_2}, a mildly increasing gain of social welfare can be gleaned as the bracket increases, though falling short of the comparable flat-rate policy. The dollar amount of social benefit, embodied by the Benefit-Cost ratio, on the other hand, trends up with the bracket level.  

\begin{flushleft}
\textbf{Example 1.5: Critical resource shortage relief}
\end{flushleft}

\begin{figure}[h!]
    \centering
    \begin{minipage}{0.45\textwidth}
        \centering
        \includegraphics[width=0.6\textwidth]{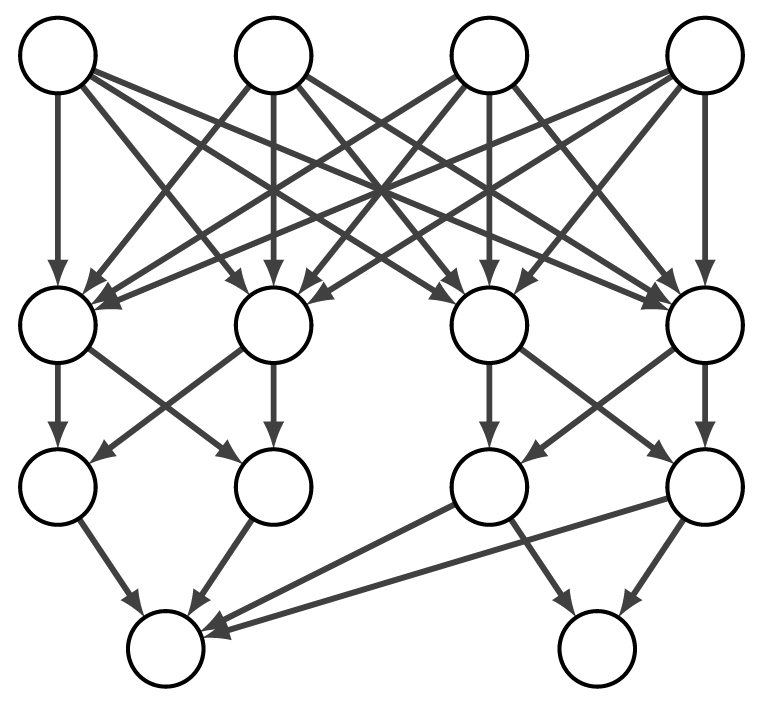} 
        \begin{center}
        (a) A latex glove shortage occurs at residential
        \end{center}        
    \end{minipage}\hfill
    \begin{minipage}{0.45\textwidth}
        \centering
        \includegraphics[width=0.6\textwidth]{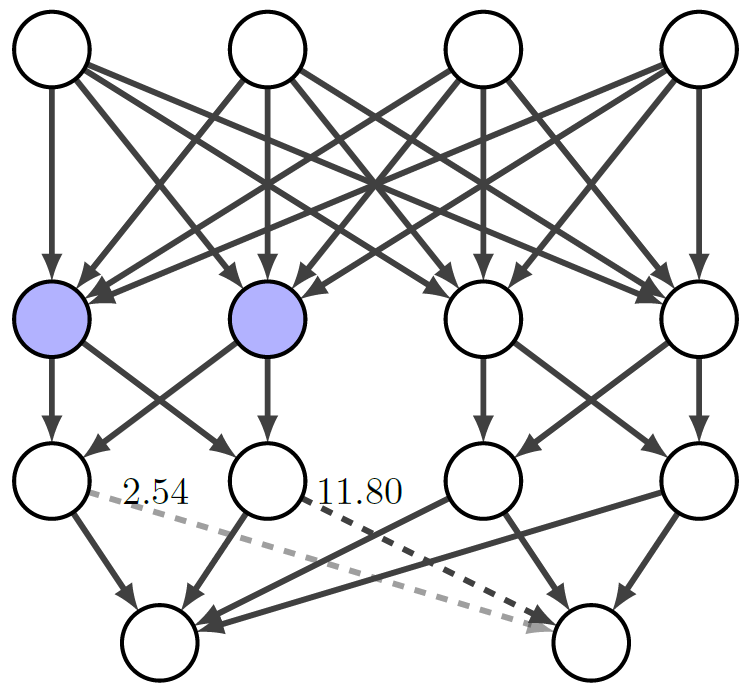} 
        \begin{center}
        (b) Shortage relieved by producer incentives
        \end{center} 
    \end{minipage}
    \caption{Critical resource shortage relief through production stimulus}
    \label{fig:eg5}
\end{figure}

In this example, we use a flat-rate incentive to relieve a latex glove shortage caused by a demand surge at the medical facilities. First, we construct a distressed supply chain in which the natural rubber emerges to be a critical shortage in its supplies at the medical facilities. In doing so, we inherit all settings from example 1.1, except the price-demand function of latex gloves at medical facilities, which is set as
	\begin{align}
		p^{1}_{31} (d_{11})= - d_{11} + 420. \nonumber 
	\end{align}	
Our algorithm returns the result including the shipments from two of the suppliers to a market, $x^{11}_{12}$ and $x^{12}_{12}$ being 0. To illustrate, Figure \ref{fig:eg5}(a) displays the topology of this disrupted supply chain, in which the residential facilities are completely cut from the supply of latex gloves. 	
To aid such circumstances, we impose a flat-rate incentive on both latex glove producers, with $\beta^1_0=50$, and re-run the model. Immediately, the previously disrupted supply can be restored. We use Figure \ref{fig:eg5}(b) to display the recovered supply chain status. Similar results of shortage relief can be achieved by imposing a flat-rate incentive on both rubber farmers.

\begin{flushleft}
\textbf{Example 2: A mixed fiscal policy in an abstract scarce resource supply chain}
\end{flushleft}

\begin{figure}[h]
		\centering
		\includegraphics[width=12cm]{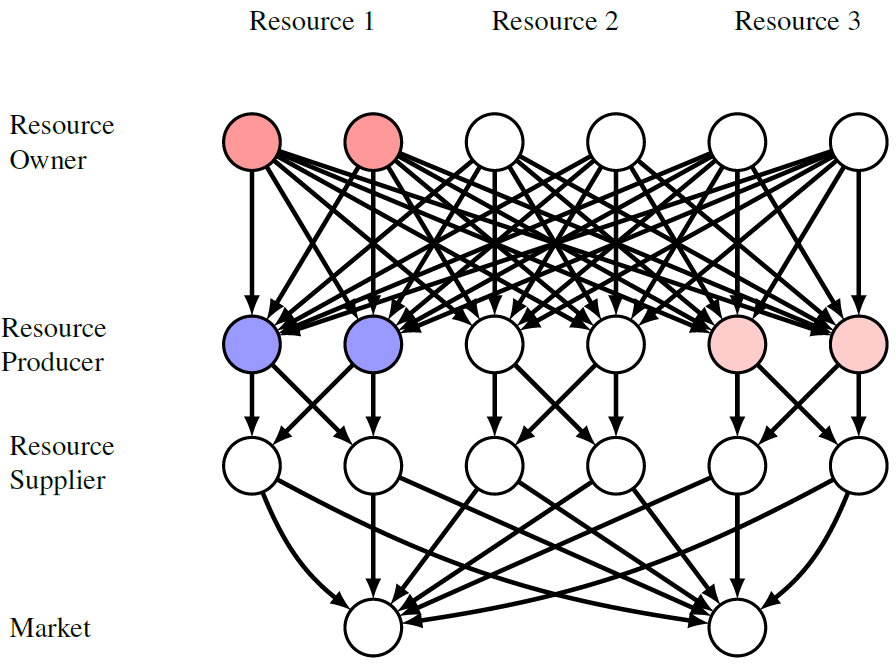}
		\caption{A resource-trio supply chain network with a mixed fiscal policy}
		\label{fig:eg6}
	\end{figure}

In many developed economies, governments tend to be concerned with levels of income inequality, and are, therefore, interested in making redistribution of societal wealth a substantial objective for economic development. Thus, it is meaningful to evaluate the utility of a mixed fiscal policy in redistributing social welfare. In this example, we construct an abstract resource-trio network to illustrate a mixed fiscal policy with a combination of incentives and taxes. In the network topology shown in Figure \ref{fig:eg6}, the red nodes indicate the firms that are being taxed whereas the blue ones incentivized. 

The cost functions are constructed, for all $i=1,...,I$ and $j=1,...,J$, with $I=J=3, N_i=M_j=K_j=2$ and $T_j=1$, as the following. 
	\begin{align}
		&f^{11}=2.5q_{11}^2+q_{11}(q_{21}+q_{31}) + 2q_{11}, \quad f^{12}=2.5q_{12}^2+q_{12}(q_{22}+q_{32}) + 2q_{12}, \nonumber \\
		& f^{13}=2.5q_{13}^2+q_{13}(q_{23}+q_{33}) + 2q_{12}, \quad f^{21}=0.5q_{21}^2+q_{21}(q_{11}+q_{31}) + 2q_{21}, \nonumber \\
    	& f^{22}=0.5q_{22}^2+q_{22}(q_{12}+q_{32}) + 2q_{22}, \quad f^{23}=0.5q_{23}^2+q_{23}(q_{13}+q_{33}) + 2q_{23}, \nonumber \\
    	&f^{31}=0.5q_{31}^2+q_{31}(q_{11}+q_{21}) + 2q_{31}, \quad f^{32}=0.5q_{32}^2+q_{32}(q_{12}+q_{22}) + 2q_{32}, \nonumber \\
    	&f^{33}=0.5q_{33}^2+q_{33}(q_{13}+q_{23}) + 2q_{33}, \quad where, \quad q_{in}=\sum^{I}_{j=1} \sum^{M_j}_{m=1} x^{in}_{jm}; \nonumber \\
		& c^{11}_{1}=0.5(x^{11}_{1})^2, \quad c^{11}_{2}=0.25(x^{11}_{2})^2, \quad c^{12}_{1}=0.5(x^{12}_{1})^2, \quad c^{12}_{2}=0.25(x^{12}_{2})^2, \nonumber\\
    & c^{21}_{1}=0.5(x^{21}_{1})^2, \quad c^{21}_{2}=0.25(x^{21}_{2})^2, \quad c^{22}_{1}=0.5(x^{22}_{1})^2, \quad c^{22}_{2}=0.25(x^{22}_{2})^2, \nonumber\\
    & c^{31}_{1}=0.5(x^{31}_{1})^2, \quad c^{31}_{2}=0.25(x^{31}_{2})^2, \quad c^{32}_{1}=0.5(x^{32}_{1})^2, \quad c^{32}_{2}=0.25(x^{32}_{2})^2; \nonumber\\
    & f^{11}=0.1(x^{11}_1 + x^{12}_1 )^2, \quad f^{12}=0.1(x^{11}_2 + x^{12}_2 )^2, \nonumber \\
    & f^{21}=0.1(x^{21}_1 + x^{22}_1 )^2, \quad f^{22}=0.1(x^{21}_2 + x^{22}_2 )^2, \nonumber \\
    & f^{31}=0.1(x^{31}_1 + x^{32}_1 )^2, \quad f^{32}=0.1(x^{31}_2 + x^{32}_2 )^2; \nonumber \\
    & c^{11}_{11}=0.5(x^{11}_{11})^2 + 3.5x^{11}_{11}, \quad c^{11}_{12}=0.5(x^{11}_{12})^2 + 3.5x^{11}_{12}, \nonumber \\
    & c^{12}_{11}=0.5(x^{12}_{11})^2 + 2x^{12}_{11}, \qquad c^{12}_{12}=0.5(x^{12}_{12})^2 + 2x^{12}_{12}, \nonumber\\
    & c^{21}_{11}=0.4(x^{21}_{11})^2 + 3.5x^{21}_{11}, \quad c^{21}_{12}=0.4(x^{21}_{12})^2 + 3.5x^{21}_{12}, \nonumber\\
    & c^{22}_{11}=0.4(x^{22}_{11})^2 + 2x^{22}_{11}, \qquad c^{22}_{12}=0.4(x^{22}_{12})^2 + 2x^{22}_{12}, \nonumber\\
    & c^{31}_{11}=0.45(x^{31}_{11})^2 + 3.5x^{31}_{11}, \quad c^{31}_{12}=0.45(x^{31}_{12})^2 + 3.5x^{31}_{12}, \nonumber\\
    & c^{32}_{11}=0.45(x^{32}_{11})^2 + 2x^{32}_{11}, \qquad c^{32}_{12}=0.45(x^{32}_{12})^2 + 2x^{32}_{12}; \nonumber\\
    & \hat{c}^{js}_{tk}(x^{js}_{tk}) =0.1x^{js}_{tk}.  \nonumber
	\end{align}
	
All other costs are set to zero. The price-demand functions at the markets are 
	\begin{align}
		p^{j}_{3k} (d_{jk})= - d_{jk} + 300,\quad \forall j,k.\nonumber 
	\end{align}	

Similar to example 1-5, we set the production conversion rates $\psi^{in}_{jm}=0.9$, the market resource commodity weights $w_{11}=0.5, w_{12}=0.5$, the parameters concerning the resource capacities, $U_i$, sufficiently large. In contrast to previous examples, here we set the step-size $\varphi$ to be $10^{-5}$, the convergence tolerance $\epsilon$ to be $6\times10^{-4}$ for this example. 

To examine the efficacy of this mixed fiscal policy, we first establish the equilibrium of the benchmark scenario, i.e., the setting without such policy. As such, we present the results in Table \ref{table:eg6}. With the ex-ante knowledge that the resource owners capture most of the supply chain profit, we then impose such a policy in which the producers of resource 1 are given a flat-rate incentive of $\beta^1_0=12$, whereas the owners of resource 1 and producers of resource 3 are charged a flat-rate tax of $\alpha^1_0=-10$ and $\beta^1_0=-2$, respectively. 

The projection method takes approximately 120 seconds for this problem of a total of 99 variables to converge to the preset tolerance. We include all equilibrium results in the supplemental file while displaying only the profit-related outcome in Table \ref{table:eg6} again. It is worth pointing out that the "net incentive" is the total taxes collected net of the total incentive disbursed by the government. 

\begin{longtable}{  llll|llll  p{13cm} } 
\caption{Profit and welfare results of example 2\label{table:eg6}}\\
\hline
Profit   &              & Benchmark & With policy & Welfare           &                 & Benchmark & With policy \\\hline
Owner    & $\pi_{11}$ & 2573.50   & 2359.58     & Consumer          & $CS_{11}$       & 104.26    & 141.64      \\
Owner    & $\pi_{12}$ & 2573.50   & 2700.43     & Consumer          & $CS_{12}$       & 120.11    & 97.84       \\
Owner    & $\pi_{21}$ & 2573.50   & 2700.43     & Consumer          & $CS_{21}$       & 107.24    & 77.30       \\
Owner    & $\pi_{22}$ & 2573.50   & 2359.58     & Consumer          & $CS_{22}$       & 104.26    & 141.64      \\
Owner    & $\pi_{31}$ & 2573.50   & 2700.43     & Consumer          & $CS_{31}$       & 120.11    & 97.84       \\
Owner    & $\pi_{32}$ & 2573.50   & 2700.43     & Consumer          & $CS_{32}$       & 107.24    & 77.30       \\
Producer & $\pi_{11}$ & 248.25    & 340.20      & Tot. owner        & $\pi_{total}$ & 15441.00  & 15520.90    \\
Producer & $\pi_{12}$ & 433.20    & 345.63      & Tot. producer     & $\pi_{total}$ & 1977.61   & 1963.28     \\
Producer & $\pi_{21}$ & 307.36    & 295.82      & Tot. supplier     & $\pi_{total}$ & 1825.59   & 667.13      \\
Producer & $\pi_{22}$ & 248.25    & 340.20      & Tot. consumer     & $CS_{total}$    & 663.21    & 633.57      \\
Producer & $\pi_{31}$ & 433.20    & 345.63      & Soc. welfare      & $SW$            & 19907.42  & 18784.87    \\
Producer & $\pi_{32}$ & 307.36    & 295.82      & Net Incentive    &                 & 0.00      & 99.15       \\
Supplier & $\pi_{11}$ & 323.04    & 175.60      & Soc. welfare gain & $\Delta SW$     & 0.00      & -1122.55    \\
Supplier & $\pi_{12}$ & 302.42    & 60.78       & Benefit-to-cost   & BC              & -         & -11.32      \\
Supplier & $\pi_{21}$ & 255.43    & 95.10       &                   &                 &           &             \\
Supplier & $\pi_{22}$ & 319.75    & 182.51      &                   &                 &           &             \\
Supplier & $\pi_{31}$ & 291.83    & 58.54       &                   &                 &           &             \\
Supplier & $\pi_{32}$ & 333.12    & 94.60       &                   &                 &           &            
\\\hline
\end{longtable}

\begin{figure}[h!]
		\centering
		\includegraphics[width=\textwidth]{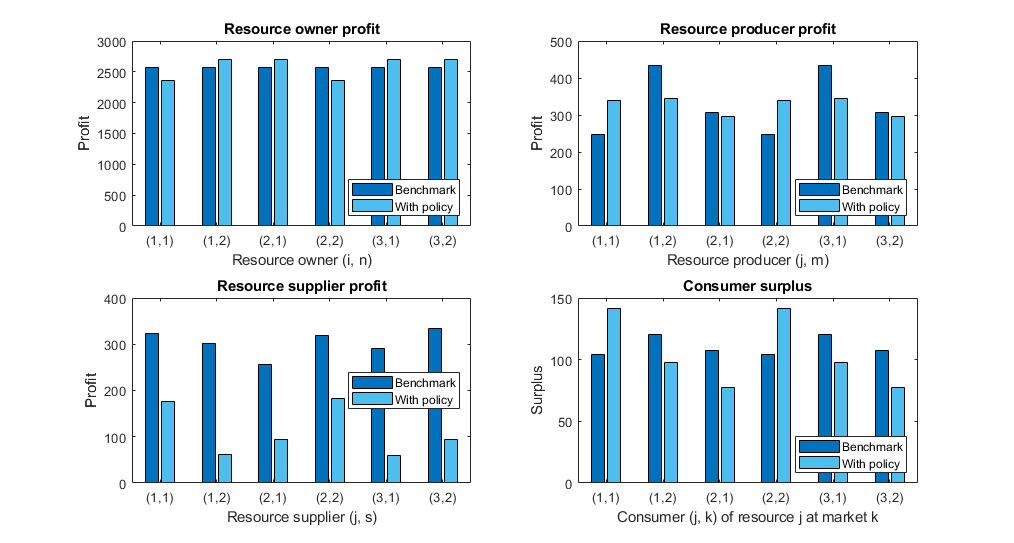}
		\caption{Profits and welfare under a mixed fiscal policy}
		\label{fig:eg6_1}
	\end{figure}

In Figure \ref{fig:eg6_1}, we observe that under the mixed fiscal policy, the directly affected firms, i.e., the top two tiers of the network, do not consistently respond to the intended policy elements, as there are both increases and decreases of profit on both the incentivized and taxed firms. On the bottom two tiers of the network, the supplier profits drop unanimously, whereas the distribution of consumer surplus across the markets becomes more uneven. 

	\begin{figure}[h!]
		\centering
		\includegraphics[width=10cm]{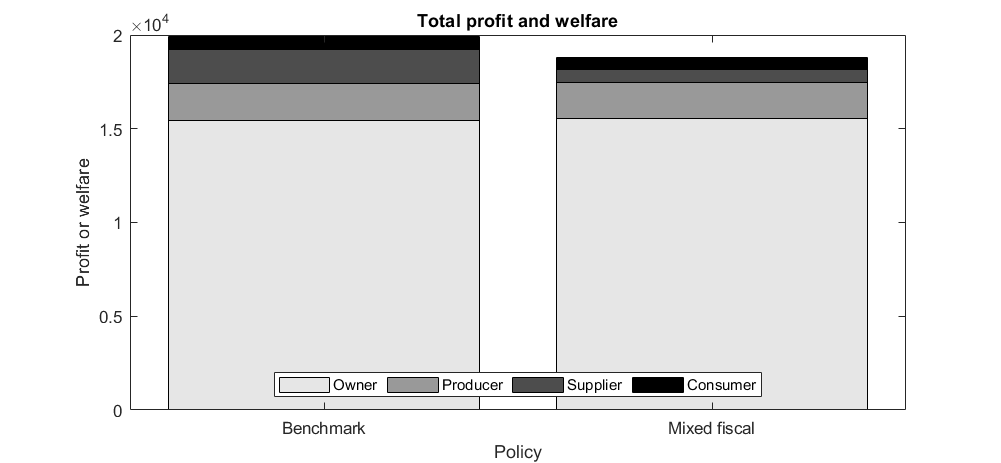}
		\caption{Total profits and welfare under a mixed fiscal policy}
		\label{fig:eg6_2}
	\end{figure}
Overall, with a positive net incentive, \$99.15, administered to the network, we observe, however, that social welfare gains a negative amount, owing to the significant decrease in supplier profits. See Table \ref{table:eg6} and Figure \ref{fig:eg6_2}. 

\section{Managerial implications}

Government plays a key role in maintaining a balance between the firm's profitability, consumer's well-being, supply chain's overall health, and society's sustainable growth. Yet in the face of widespread resource shortages that stifle the economic growth in supply chains, balancing the tradeoffs between the goals has been enormously challenging for governments. The proposed cross-sector multi-product scarce resource supply chain network model can serve as a support system for both the supply chain practitioners and governments to make operational and strategic planning decisions. With a particular focus on the practice that can be adopted by the government to help address the insurgent scarcity and achieve overall sustainability, we herein instill the following managerial implications.

\begin{enumerate}

\item From the viewpoint of designing a stimulative fiscal policy, the legislator will likely face a caveat between the choice of a flat-rate policy and a regressive one. Our analysis shows that the flat-rate incentive generally performs more effectively than its regressive counterpart when administered to the resource owners in stimulating supply chain welfare. This result is in general agreement with a dual study by \citet{Yu2018}, in which they pointed out the strength of flat-rate tax when imposed to discourage adverse environmental activities. This result is also supported by the advantages of approximately linear tax in maintaining welfare \citep{mirrlees1971exploration}. Furthermore, as a case in point, we also find that if the legislator's focus of policy design shifts to the efficiency of it, i.e., the social-welfare gain on every dollar of fiscal investment, then a carefully designed two-bracket regressive incentive can be more advantageous in policy efficiency than its flat-rate counterpart. That is, with a large enough ``bracket'', the legislator may be able to devise a regressive incentive scheme that can achieve a higher Benefit-Cost ratio, and almost as high of a net welfare gain as the flat-rate scheme. Though delicate, the value of such a bracket can be practically narrowed down by determining the largest effective bracket, followed by determining the largest achievable net welfare gain. 

\item For the choice of incentive recipients in a supply chain, i.e., the resource owners or the resource producers, the decision-makers should anticipate a tradeoff. When the incentives are administered to the owners, both the social welfare gain and the policy efficiency will remain at a relatively high level. When the incentives are administered to the producers, the supplier's profit will enjoy the most gains, whereas the market price of the affected products will be reduced significantly. In practice, the market price of a scarce resource not only influences the consumer's behaviors, but also serves as a key indicator for the broader economy. Under a supply chain disruption, we suggest that the governments first administer an appropriate amount of incentives to the producers so that the commodity prices can be reduced at the market level in short term. We note that, at the same time, any commodities experiencing shortage at the market level may also be relieved by the same incentives. Once the cause for the disruption has subsided and the supply chain's performance measures stabilized, governments can then reelect the appropriate supply chain tier(s) as the new recipients of the next round of incentives, depending on the economic objectives and legislative priorities.

\item “Income inequality” is believed to hinder economic growth. Using fiscal policy to reduce income gaps has become the goal of many advanced economies \citep{coady2012income}. In our observation, the large profit difference across the supply chain tiers entices a growth-minded government to pursue a redistributive strategy for social surplus. It has been found, however, that the redistribution of welfare among resource owners and producers via a mixed fiscal policy may result in a net loss of welfare. While we acknowledge that such a finding resembles the rhetoric of the opposition to the dominant social policy notion that the resources generated by economic growth should be redistributed to fund social programs \citep{midgley1999growth}, it is pertinent to note that our analytical takeaway is derived from a microeconomic framework with standard assumptions. Nonetheless, we note further that the welfare loss caused by our experimental mixed fiscal policy has been previously associated with the elasticity of demand from the classic oligopoly theories \citep{worcester1975monopoly}. In particular, it is also widely acknowledged in literature that taxation can be ineffective in reducing wealth inequalities as opposed to what conventional wisdom would have anticipated \citep{mirrlees1971exploration}. As is demonstrated before, our proposed fiscal policy can effectively relieve resource shortages and stimulate growth in supply chains. But for the governments or practitioners who oversee the supply chain grand strategies in a post-crisis stage, we caution that the well-intended fiscal interventionist may derail the overall economic sustainability of society. We advise that the use of a mixed fiscal policy in supply chains should follow the principle of configuring a mild or less redistributive policy, as is recognized in \citet{coady2012income}.

\item Finally, we note an interesting link between resource capacity and welfare within a competitive supply chain. In related literature, \citet{chen2008supply} imposed firm-wise resource capacity in their supply chain and found that capacity limit restricts welfare. In both \citet{nagurney2019strict} and \citet{hu2013equilibrium}, though the shared capacity limit was modeled, neither study associated such a limit with its impact on social welfare. The current study, however, proffers further results. We find that the capacity limit of a given type of scarce (more often, natural) resource does not strictly curtail social welfare. Rather, an appropriate level of the limit can even benefit the social outcome. In practice, the ownership and the right to use a natural resource as a shared public good is often governed by its respective common laws or local policies\footnote{Public goods has been a widely discussed topic in studies of economics and law. See \citet{reaume1988individuals}, \citet{holcombe1997theory} and the reference therein.}. Thus, if a natural resource of critical importance belongs to a government’s jurisdiction, it is advisable for the government to legislate and impose a mild restriction on the usage of such natural resources. The restriction, if properly selected, will not only preserve the quantity of the shared resource but also create a “sweet spot” that induces higher social welfare.  

\end{enumerate}

Admittedly, the above managerial implications are based on our stylized numerical experiments and thus, are limited to the extent of the characteristics of the network, e.g., the competitive nature of the nodes, the substitutability of the flows, the mix of policies, etc. Moreover, because of the number of features incorporated in our model, it is probable that in the presence of multiple features, their interactions could result in more profound managerial implications than what we have uncovered. Nonetheless, practitioners and decision-makers should carefully verify and validate the premises of the model before expanding the aforementioned insights.

\section{Conclusions}

Prima facie, the munificence of scarce resources is akin to the sustainability and growth of individual firms, societies, and the flourishing of humanity. Yet, in this age of intensifying societal changes, shocks, crises, and inter-connectivity, the conflict and competition for scarce resources and products have become more fierce. Fiscal policy remains the most common governmental policy instrument to relieve the shortage of supplies and stimulate economies. In this paper, our contributions to the literature on scarce resources and supply chain networks include the following. 

We construct the first general decentralized cross-sector scarce resource SCNE model with a unifying supply-side fiscal policy. We provide a rigorous VI formulation for the governing equilibrium conditions of the network model. Such a substitute network provides a versatile tool for the evaluation of profit, welfare, policy instruments, cost structure, transportation, conservation, competition, and interdependence of resources throughout the supply chains. The generality of our model also allows for a variety of extensions, i.e., dynamics, stochastic features, multi-criteria decision-making, disequilibrium behaviors, etc, to be furnished. 

Second, our model is also a general Nash equilibrium problem. We formulate the GNE of the network model in VI. There are only a few GNEP studies that incorporate fiscal policy in the SCNE literature. The utility of this model is not limited to the scarce resource supply chains, but also eligible to any resource commodity that pertains to the aforementioned characteristics of scarce resources. 

Third, from a technical aspect, we introduce a recently uncovered approach to characterize the network equilibrium by adopting a novel set of theoretical properties, including $\lambda_{min}$. To the best of our knowledge in the supply chain network literature, such a means of characterization for the uniqueness property of network equilibrium has yet to appear heretofore. 

Lastly, we furnish the model with numerical studies and extract managerial insights that provide governments, resource owners, and firms useful advice in expansion, cost restructuring, resource conservation, competition/collaboration strategy, shortage handling, and post-crisis stimulation. In particular, we provide guidance on supply-side policy design and administration in relieving and stimulating the PPE shortage caused by the COVID-19 global pandemic. Our findings also enrich the political discussion on public resource legislation, income inequality, and sustainable development. We anticipate that the extension of this model can shed light on the stimulation and relief effort on vaccine distribution and economic recovery.

%%%%%%%%%%%%%%%%%%%%%%% REFERENCE %%%%%%%%%%%%%%%%%%%%%%%%

\singlespacing		% from setspace package
%\bibliographystyle{apalike}
%\bibliography{/Users/Xiaowei Hu/Documents/library.bib}
\bibliography{library_sr}
\onehalfspacing		% from setspace package

\pagebreak

%%%%%%%%%%%%%%%%%%%%%%% APPENDICES %%%%%%%%%%%%%%%%%%%%%%%%

\appendix

\begin{center}
{\LARGE \textbf{APPENDIX}}
\end{center}

\section{Proof of Theorem 1}\label{appendix:a}

\textbf{Proof:}	First, we prove that an equilibrium according to Definition 1 coincides with the solution of VI (\ref{eqn:60}). The summation of (\ref{eqn:owneropt2}), (\ref{eqn:produceropt2}), (\ref{eqn:supplieropt2}), and (\ref{eqn:marketopt}), after algebraic simplifications, yields (\ref{eqn:60}). 

Next, we prove the converse, that is, a solution to the VI (\ref{eqn:60}) satisfies the sum of conditions (\ref{eqn:owneropt2}), (\ref{eqn:produceropt2}), (\ref{eqn:supplieropt2}), and (\ref{eqn:marketopt}), and thereby, is a cross-sector multi-product scarce resource SCNE pattern, in accordance with Definition 1. 

In (\ref{eqn:60}), we begin by adding the term $\sum_{j=1}^{I} \sum_{m=1}^{M_j}(p^{in*}_{0jm} - p^{in*}_{0jm}) $ to the first summand expression over $i$ and $n$, $ \sum_{s=1}^{S_j}(p^{jm*}_{1s} - p^{jm*}_{1s}) $ to the third summand expression over $j$ and $m$, and lastly, $p^{js*}_{2tk} - p^{js*}_{2tk} $ to the fifth summand expression over $i$, $s$, $t$, and $k$. Since these terms are all equal to zero, (\ref{eqn:60}) holds true. Hence, we obtain the following inequality: 

  \begin{align}\label{eqn:65}
	\sum_{i=1}^{I} \sum_{n=1}^{N_i} \sum_{j=1}^{I} \sum_{m=1}^{M_j} \Bigg[ \frac{\partial f^{in}(x^{in*})}{\partial x^{in}_{jm}} + \frac{\partial f^{jm}(x^*_{jm})}{\partial x^{in}_{jm}} + \frac{\partial c^{in}_{jm}(x^{in*}_{jm})}{\partial x^{in}_{jm}} - \frac{\partial \alpha^i_0(x^{in*}_{jm})}{\partial x^{in}_{jm}}   \nonumber \\
	+ \lambda^{0*}_i - \psi^{in}_{jm} \lambda^{1*}_{jm} + \sum_{g=1}^{G}\mu^{0*}_{ing} + (p^{in*}_{0jm} - p^{in*}_{0jm}) \Bigg] & \times (x^{in}_{jm} - x^{in*}_{jm}) \nonumber \\
	+ \sum_{i=1}^{I} \sum_{n=1}^{N_i} \sum_{g=1}^{G} \bigg[ - \frac{\partial \alpha^i_g(\delta^{in*}_g)}{\partial \delta^{in}_g} - \mu^{0*}_{ing} \bigg] & \times (\delta^{in}_g - \delta^{in*}_g )  \nonumber \\
	+ \sum_{j=1}^{I} \sum_{m=1}^{M_j}  \sum_{s=1}^{S_j} \Bigg[ \dfrac{\partial f^{js}(x^{js*})}{\partial x^{jm}_{s}} + \frac{\partial c^{jm}_{s}(x^{jm*}_{s})}{\partial x^{jm}_{s}} -  \frac{\partial \beta^j_0(x^{jm*}_{s})}{\partial x^{jm}_{s}} \nonumber \\ 
	+ \lambda^{1*}_{jm} - \lambda^{2*}_{js} + \sum_{g=1}^{G} \mu^{1*}_{jmg} + (p^{jm*}_{1s} - p^{jm*}_{1s})  \Bigg] & \times (x^{jm}_{s} - x^{jm*}_{s})  \nonumber \\
	+ \sum_{j=1}^{I} \sum_{m=1}^{M_j} \sum_{g=1}^{G} \bigg[ -\frac{\partial \beta^j_g(\delta^{jm*}_g)}{\partial \delta^{jm}_g} - \mu^{1*}_{jmg} \bigg] & \times (\delta^{jm}_g - \delta^{jm*}_g )  \nonumber \\
	+ \sum_{j=I}^{I} \sum_{s=1}^{S_j} \sum_{t=1}^{T_j} \sum_{k=1}^{K} \bigg[ \frac{\partial c^{js}_{tk}(x^{js*}_{tk})}{\partial x^{js}_{tk}} + \hat{c}^{js}_{tk}(x^{js*}_{tk}) + \lambda^{2*}_{js} + (p^{js*}_{2tk} - p^{js*}_{2tk})  \bigg] & \times (x^{js}_{tk}-x^{js*}_{tk})  \nonumber \\
	- \sum^{I}_{j=1} \sum^{K}_{k=1} p^{j}_{3k} (d^*) & \times (d_{jk} -d^*_{jk})  \\
	+ \sum_{i=1}^{I} \bigg[U_i - \sum_{n=1}^{N_i} \sum_{j=1}^{I} \sum_{m=1}^{M_j} x^{in*}_{jm} \bigg] & \times  (\lambda^0_i - \lambda^{0*}_i) \nonumber \\ 
	+ \sum_{j=1}^{I} \sum_{m=1}^{M_j} \bigg[ \sum_{i=1}^{I} \sum_{n=1}^{N_i} x^{in*}_{jm}\cdot\psi^{in}_{jm} - \sum_{s=1}^{S_j} x^{jm*}_{s} \bigg] & \times (\lambda^1_{jm} - \lambda^{1*}_{jm})  \nonumber \\
	+ \sum_{j=I}^{I} \sum_{s=1}^{S_j} \bigg[ \sum_{m=1}^{M_j} x^{jm*}_{s} - \sum_{k=1}^{K} \sum_{t=1}^{T_j} x^{js*}_{tk} \bigg] & \times (\lambda^2_{js}-\lambda^{2*}_{js})  \nonumber \\
	+ \sum_{i=1}^{I} \sum_{n=1}^{N_i} \sum_{g=1}^{G} \bigg[A^i_g - \sum^{I}_{j=1} \sum^{M_j}_{m=1} x^{in*}_{jm} + \delta^{in*}_g \bigg] & \times ( \mu^0_{ing} - \mu^{0*}_{ing})  \nonumber \\
	+ \sum_{j=1}^{I} \sum_{m=1}^{M_j} \sum_{g=1}^{G} \bigg[ B^j_g - \sum^{S_j}_{s=1} x^{jm*}_{s} + \delta^{jm*}_g \bigg] & \times (\mu^1_{jmg} - \mu^{1*}_{jmg}) \geq 0,  \nonumber \\
	\forall (Q^0,Q^1,Q^2, \mathfrak{\Delta}^0, \mathfrak{\Delta}^1, d, \lambda^0, \lambda^1, \lambda^2, \mu^0,\mu^1) \in \mathcal{K}. \nonumber 
  \end{align}   

Rearranging (\ref{eqn:65}) yields:

  \begin{align}\label{eqn:66}
	\sum_{i=1}^{I} \sum_{n=1}^{N_i} \sum_{j=1}^{I} \sum_{m=1}^{M_j} \bigg[ \frac{\partial f^{in}(x^{in*})}{\partial x^{in}_{jm}} + \frac{\partial c^{in}_{jm}(x^{in*}_{jm})}{\partial x^{in}_{jm}} - p^{in*}_{0jm} - \frac{\partial \alpha^i_0(x^{in*}_{jm})}{\partial x^{in}_{jm}}  \nonumber \\
	+ \lambda^{0*}_i + \sum_{g=1}^{G}\mu^{0*}_{ing} \bigg] & \times (x^{in}_{jm} - x^{in*}_{jm}) \nonumber \\
	+ \sum_{i=1}^{I} \sum_{n=1}^{N_i} \sum_{g=1}^{G} \bigg[ - \frac{\partial \alpha^i_g(\delta^{in*}_g)}{\partial \delta^{in}_g} - \mu^{0*}_{ing} \bigg] & \times (\delta^{in}_g - \delta^{in*}_g ) \nonumber \\ 
	+ \sum_{i=1}^{I} \sum_{n=1}^{N_i} \sum_{j=1}^{I} \sum_{m=1}^{M_j} \bigg[ \frac{\partial f^{jm}(x^*_{jm})}{\partial x^{in}_{jm}} +  p^{in*}_{0jm} - \psi^{in}_{jm} \lambda^{1*}_{jm} \bigg] & \times (x^{in}_{jm}-x^{in*}_{jm}) \nonumber \\
	+ \sum_{j=1}^{I} \sum_{m=1}^{M_j} \sum_{s=1}^{S_j} \bigg[ \frac{\partial c^{jm}_{s}(x^{jm*}_{s})}{\partial x^{jm}_{s}} - p^{jm*}_{1s} - \frac{\partial \beta^j_0(x^{jm*}_{s})}{\partial x^{jm}_{s}} + \lambda^{1*}_{jm} + \sum_{g=1}^{G} \mu^{1*}_{jmg} \bigg] & \times (x^{jm}_{s} - x^{jm*}_{s}) \nonumber \\
	+ \sum_{j=1}^{I} \sum_{m=1}^{M_j} \sum_{g=1}^{G} \bigg[ -\frac{\partial \beta^j_g(\delta^{jm*}_g)}{\partial \delta^{jm}_g} - \mu^{1*}_{jmg} \bigg] & \times (\delta^{jm}_g - \delta^{jm*}_g ) \nonumber \\
	+ \sum_{j=I}^{I} \sum_{s=1}^{S_j} \sum_{t=1}^{T_j} \sum_{k=1}^{K} \bigg[ \frac{\partial c^{js}_{tk}(x^{js*}_{tk})}{\partial x^{js}_{tk}} - p^{js*}_{2tk} + \lambda^{2*}_{js} \bigg] & \times (x^{js}_{tk}-x^{js*}_{tk}) \nonumber \\
	+ \sum_{j=1}^{I} \sum_{m=1}^{M_j} \sum_{s=1}^{S_j} \bigg[ \dfrac{\partial f^{js}(x^{js*})}{\partial x^{jm}_{s}} + p^{jm*}_{1s} - \lambda^{2*}_{js} \bigg] & \times (x^{jm}_{s}-x^{jm*}_{s}) \nonumber \\
   + \sum_{j=1}^{I} \sum_{s=1}^{S_j} \sum_{t=1}^{T_j} \sum_{k=1}^{K} [p^{js*}_{2tk} + \hat{c}^{js}_{tk}(x^{js*}_{tk})] & \times (x^{js}_{tk}-x^{js*}_{tk}) \nonumber \\
    - \sum^{I}_{j=1} \sum^{K}_{k=1} p^{j}_{3k} (d^*) & \times (d_{jk} -d^*_{jk}) \\
   + \sum_{i=1}^{I} \bigg[U_i - \sum_{n=1}^{N_i} \sum_{j=1}^{I} \sum_{m=1}^{M_j} x^{in*}_{jm} \bigg] & \times  (\lambda^0_i - \lambda^{0*}_i) \nonumber \\ 
	+ \sum_{j=1}^{I} \sum_{m=1}^{M_j} \bigg[ \sum_{i=1}^{I} \sum_{n=1}^{N_i} x^{in*}_{jm}\cdot\psi^{in}_{jm} - \sum_{s=1}^{S_j} x^{jm*}_{s} \bigg] & \times (\lambda^1_{jm} - \lambda^{1*}_{jm}) \nonumber \\
	+ \sum_{j=I}^{I} \sum_{s=1}^{S_j} \bigg[ \sum_{m=1}^{M_j} x^{jm*}_{s} - \sum_{k=1}^{K} \sum_{t=1}^{T_j} x^{js*}_{tk} \bigg] & \times (\lambda^2_{js}-\lambda^{2*}_{js}) \nonumber \\
	+ \sum_{i=1}^{I} \sum_{n=1}^{N_i} \sum_{g=1}^{G} \bigg[A^i_g - \sum^{I}_{j=1} \sum^{M_j}_{m=1} x^{in*}_{jm} + \delta^{in*}_g \bigg] & \times ( \mu^0_{ing} - \mu^{0*}_{ing}) \nonumber \\
	+ \sum_{j=1}^{I} \sum_{m=1}^{M_j} \sum_{g=1}^{G} \bigg[ B^j_g - \sum^{S_j}_{s=1} x^{jm*}_{s} + \delta^{jm*}_g \bigg] & \times (\mu^1_{jmg} - \mu^{1*}_{jmg}) \geq 0, \nonumber \\
	\forall (Q^0,Q^1,Q^2, \mathfrak{\Delta}^0, \mathfrak{\Delta}^1, d, \lambda^0, \lambda^1, \lambda^2, \mu^0,\mu^1) \in \mathcal{K}.  \nonumber
  \end{align}

Clearly, (\ref{eqn:66}) is the sum of the optimality condition (\ref{eqn:owneropt2}), (\ref{eqn:produceropt2}), (\ref{eqn:supplieropt2}), and (\ref{eqn:marketopt}), and thereby, is, according to Definition \ref{def:def1}, a cross-sector multi-product scarce resource SCNE pattern.	$\square$

\end{document}